\documentclass[manuscript]{aastex}
\usepackage[top=1in, bottom=1in, left=1.0in, right=1in]{geometry}
\usepackage{graphics,natbib,lscape}
\usepackage{multirow}
\usepackage{longtable}

\shorttitle{Binary Frequencies in Globular Clusters}
\shortauthors{Ji \& Bregman}

\begin{document}

\title{Binary Frequencies in a Sample of Globular Clusters. II. Sample Analysis and Comparison to Models}

\author{Jun Ji\altaffilmark{1} and Joel N. Bregman}

\affil{Department of Astronomy, University of Michigan, Ann Arbor, MI 48109}
\email{jijunatumich@gmail.com, jbregman@umich.edu}

\altaffiltext{1}{Princeton Consultants, 2 Research Way  Princeton, NJ 08540}

\begin{abstract}

Binary stars are predicted to have an important role in the evolution of globular clusters, 
so we obtained binary fractions for 35 globular clusters that were imaged in the F606W 
and F814W with the ACS on the Hubble Space Telescope.  When compared to the values of 
prior efforts \citep{sollima07, milone12}, we find significant discrepancies, despite 
each group correcting for contamination effects and having performed the appropriate reliability tests.
The most reliable binary fractions are obtained when restricting the binary fraction to $q \ge 0.5$.  
Our analysis indicates that the range of the binary fractions is nearly an order of 
magnitude for the lowest dynamical ages, suggesting that there is a broad distribution in 
the binary fraction at globular cluster formation.  Dynamical effects also appears to decrease 
the core binary fractions by a factor of two over a Hubble time, but this is a weak relationship.  
We confirm a correlation from previous work that the binary fraction within the core radius decreases with cluster age, indicating that younger clusters formed with higher binary fractions.  
The strong radial gradient in the binary fraction with cluster radius appears to be a 
consequence of dynamical interactions.  It is often not present in dynamically young clusters 
but nearly always present in dynamically old clusters.

\end{abstract}


\section{Introduction}
Globular clusters have dynamical properties that distinguish them from other stellar systems and allow one to test dynamical models.  They can be very long-lived, with typical ages for Galactic globular clusters of 7-13 Gyr \citep{salaris02}, yet they are dynamically active in that their dynamical relaxation times are significantly less than a Hubble time \citep{hut92a}.  For the well-observed Galactic globular clusters, the ratio of the age to the relaxation time (at the half-light radius) lies in the range 2-40 (a median of 7-8), representing clusters that display modest dynamical evolution to those that have undergone core collapse.

Binary stars are expected to play a central role in the dynamical evolution of globular clusters through the process of binary burning.  In this process, the dynamical interactions of binaries with other stars or binaries both hardens the binary and adds kinetic energy to the interacting star (other binary), thereby slowing the contraction of the cluster, especially in the core region.  Therefore, there are predicted relationships between the globular cluster binary properties and its dynamical state.  Some of these predictions can be tested by measuring the binary fraction of globular cluster stars between systems in different dynamical states and as a function of radius within individual clusters.

The process of measuring binary fractions with globular cluster color-magnitude diagrams relies on accurate magnitude determinations of $\sim 10^4$ Main Sequence stars in crowded environments.  
This approach is very challenging for ground-based observers, but it became feasible with the high angular resolution capabilities of the Hubble Space Telescope \citep{sollima07, milone12}.  
HST has been used to observe globular clusters with the WFPC2 and more recently with the WFC/ACS as part of a Treasury project.  
The results of these efforts on the topic of binary fractions were recently completed using these samples \citep{sollima07, ji2011, milone12, ji2013}, with a variety of interesting results.  
The methods used by \citet{ji2011} are published in \citet{ji2013} (henceforth, Paper I).  
The present paper analyzes the trends of binary fraction with cluster properties and with radius.  
Both methods and analysis are contained within the thorough analysis of \citet{milone12}, and while there are a many similarities in the approaches of both efforts, there are a few differences, both technically and in sample selection.  
Several of the results by \citet{sollima07} and \citet{milone12} are confirmed by us, but there are some important differences that we highlight in this paper.

\section{Binary Fraction Determinations within the Sample}

The sample, which was frozen in 2010, derives from the Galactic globular cluster list \citep{harris96} for which there were sufficiently long WFPC2 or ACS observations (e.g., snapshot observations were excluded).  
We avoided clusters with high extinctions (E(B-V) $>$ 0.4) to avoid the challenges associated with the variations in the extinction across the globular cluster (Paper I).  \citet{milone12} included more globular clusters with high extinction, for which they developed an approach to correct for small-scale extinction issues.  They also included other data sets, some taken after our cutoff date, so our sample contains 35 clusters \citep{ji2013} while theirs contains 59.  The earlier survey by \citet{sollima07}
examines 13 lower surface brightness globular clusters.

Briefly, the color magnitude method of determining the Main Sequence binary fraction relies on the fact that when a second star is within the resolution element, the combination departs from the Main Sequence along a well-defined locus in color and magnitude.  This approach necessitates that the measurement error is small compared to the deviation from the Main Sequence of a typical binary, such as one with a mass ratio of 1:2.  Practically, this is best accomplished with a magnitude error near 0.01 mag, which can be challenging in crowded fields, but this challenge has been met through the development of point source extraction software that includes detailed treatments of the point spread function (psf) in such situations \citep{dolphin00, anderson08}.  

We calculate three measures of the binary fraction, two of which are non-parametric.  One approach is only to count stars with a mass fraction greater than q = 0.5 and form a binary fraction, f$_b$(high q).  Another method is to fit a Gaussian distribution of the star density perpendicular to the straightened Main Sequence (as a function of color), mostly using the blue side.  The fit is extended to the red side and subtracted from the total distribution, having made corrections for contamination by false binaries and field stars.  The remaining stars are assumed to be binaries whose mass ratios extend below q = 0.5, typically reaching 0.3; the binary fraction is given by f$_b$(count).  The final method is parametric in that we produce a best-fit using a binary fraction with a power-law dependence on mass (for q $\geq$ 0.3), along with the Gaussian described above and the contamination corrections. This resulting binary fraction is  denoted as f$_b$(fit).

Contamination is a critical issue because the chance superposition of two unrelated stars is indistinguishable from a true binary.  This problem is solved by simulations that lead to estimates for the number of chance superpositions as a function of magnitude and position in the cluster.

Another type of contamination comes from foreground and background stars.  This is not a serious problem except at low latitudes, where we correct for it through Galactic models that predict the contributions from such stars \citep{Robin03}.  However, when the field is dense with foreground and background stars, these corrections are not always of sufficient precision.  A correction that is too large will produce a negative star density in the region that we are counting binaries, leading to a non-physical negative binary fraction.  This contamination most affects the $q>0.5$ method and we have two globular clusters where this occurred.

Multiple stellar populations have been discovered in an increasing number of globular clusters \citep{grat2012} and it is suspected that two generations of stars exist in globular clusters, formed typically within $10^8$ years of each other, a period of time far less than the age of these systems.  The evidence for the two (or more) populations comes from color magnitude diagrams that use ultraviolet and optical colors, and with some spectral data \citep{piot2012, mone2013, piotto15}.  For example, in NGC 6352 \citep{piotto15}, the width in color of the main sequence below the turnoff is 0.116 mag in $m_{F275W}-m_{F336W}$, which is wider than the color range used for binary fitting.  If this were the case for the CMD constructed with F606W and F814W data, determining binary fractions would be particularly challenging, if not impossible.  However, the spread of the main sequence is significantly less when using  $m_{F606W}-m_{F814W}$, because metallicity variations shift the relevant part of the main sequence in a vector nearly aligned with the main sequence (Paper I).  Therefore, multiple populations can have a nearly degenerate color sequence on the lower main sequence.  For NGC 6352, the distribution of the stars about the modal color value is largely fit by us with a Gaussian with $\sigma \approx 0.02$ mag, well less than the $\approx 0.1$ mag range used for binary determinations.  For NGC 6753 \citep{dotter15}, populations A and C, which differ mainly in their $\alpha$ enhancement and less so in the Fe abundance, are distinct in UV colors and are successfully modeled with stellar atmosphere and isochrone codes.  However, when using $m_{F606W}-m_{F814W}$, the isochrones of the two populations are degenerate, based upon the stellar atmosphere models (their figure 11).

We can quantify the spread a bit better for the  $m_{F606W}-m_{F814W}$ colors in the lower main sequence ($m_{F814W} \approx 20$ in their Figure 6), based on the models used for NGC 7089 \citep{milone15}; this is an unusual globular cluster not included in our study.  For fixed helium, age, and iron abundance (Y = 0.248, 12 Gyr, [Fe/H] = -1.0), a change in the alpha abundances from [$\alpha$/Fe] = 0.2 to 0.4 leads to a color shift in the main sequence of 0.008 mag.  The large difference in the metallicity from [Fe/H] = -1.7  (Y = 0.246, 13 Gyr, [$\alpha$/Fe] = 0.4) to a slightly younger but more metal rich population with [Fe/H] = -1.0  (Y = 0.248, 12 Gyr, [$\alpha$/Fe] = 0.4) leads to a shift of 0.029 mag.  Such large metallicity differences are uncommon and more typical case is NGC 6656 \citep{mari2009}, where the authors consider two populations with [Fe/H] = -1.82 and -1.68.  This difference leads to a measurable broadening near the turnoff, but further down the main sequence, we calculate that the broadening is 0.008 mag for the lower main sequence in the $m_{F606W}-m_{F814W}$ colors. Populations with enhanced He abundances are considered for NGC 7089 by \citet{milone15}, which lead to similar shifts, although these populations do not seem to account for much of the population on their lower main sequence or they fit poorly near the turnoff region (their Figure 6).  Also, there is no direct (spectroscopic) proof that the He abundance is enhanced by the large amounts they considered (Y = 0.33).

Multiple stellar populations certainly can add to the dispersion of the lower main sequence, but not necessarily to a degree that invalidates the determination of binary fractions.  A measure of the contamination is the breadth of the Gaussian used to fit the central part of the aligned CMD in color space.  For cases where the Gaussian fit has  $\sigma > 0.03$, the main sequence is sufficiently wide to invalidate the CMD approach to determining the binary fraction, as in the case of NGC 2808 \citep{Piotto07}; other objects were excluded from this study at the outset. However, when the main sequence is narrow, multiple main sequences are sufficiently degenerate to allow an estimation of the binary fraction.  The determination of $f_b(q>0.5)$ is probably the most reliable, as it does not depend on the main sequence having a symmetric structure.  To estimate the binary fraction to smaller values of $q$ (to about 0.3) requires a fitting procedure that makes use of the entire distribution, but depends upon the assumption that most of an individual color distribution can be represented by Gaussians.  While such fits lead to acceptable values of ${\chi}^2$, small intrinsic asymmetries 2-3$\sigma$ from the center of the distribution can lead to overestimates of the binary fraction.  Many of the fitted binary fractions, $f_b(fit)$ are consistent with an extension of $f_b(q~>~0.5)$ to $q \approx 0.3$, yet it is difficult to predict a-priori which are reliable and which are not.  Therefore, we advise investigators to use $f_b(fit)$ with caution, with $f_b(q~>~0.5)$ being more reliable.  Addition discussion on these topics is found in Paper I.

\subsection{Binary Fractions Within The Whole Field of View}

Analysis of the whole field of view statistically gives the most reliable results, as it includes the most number of stars.  This allows greater precision when comparing the binary fractions from different techniques.  However, there are shortcomings to this approach.  Due to the variation in distances, the sampling of each cluster extends to different radii.  
For example, the half light radius of NGC 104 is not contained within the ACS field.  
This issue would not be relevant except that there is a gradient in the binary fraction as a function of radius.  
We do not include the region near the CCD edge and between the gap of the two chips because the photometry is not sufficiently accurate.

In Table ~\ref{tab:whole}, we list the analysis results using different methods for the whole field of view. In this table, we give the region size in terms of their half mass radii in Column 2. In Column 3, 4, and 5, we list the binary fractions with 1 $\sigma$ errors for the high mass-ratio ($q>0.5$) method, the star counting method, and the fitting method, respectively. Column 6 and 7 are the fitted parameters for the third method, where power $x$ is the power index of the power-law function for the binary mass-ratio distribution, and $q_{min}$ is the minimum binary mass ratio we can get. Column 8 gives the $\chi^2$ and degrees of freedom for the fitting method. The last column gives the binary fraction quality flags. They indicate the main error sources. Only the results including one quality flag can be used, such as g (good estimate), d (dense core), and n (small number of stars). Results including flag f (field stars) or e (large photometric errors) should be used with caution, as the contaminations are asymmetric and with large uncertainties (some fractions become negative values due to the asymmetric distribution of field stars).

\begin{figure}
\plottwo {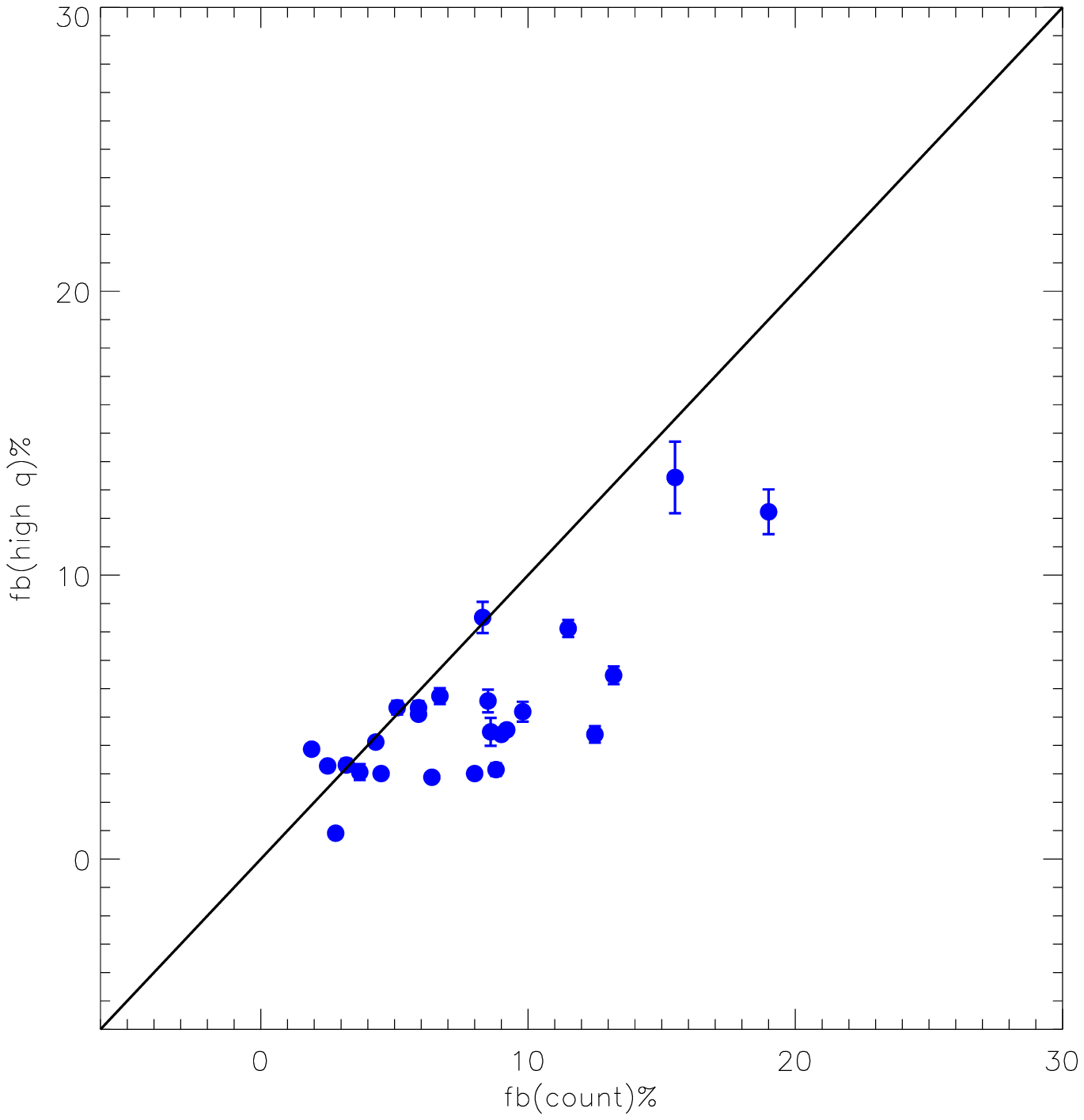} {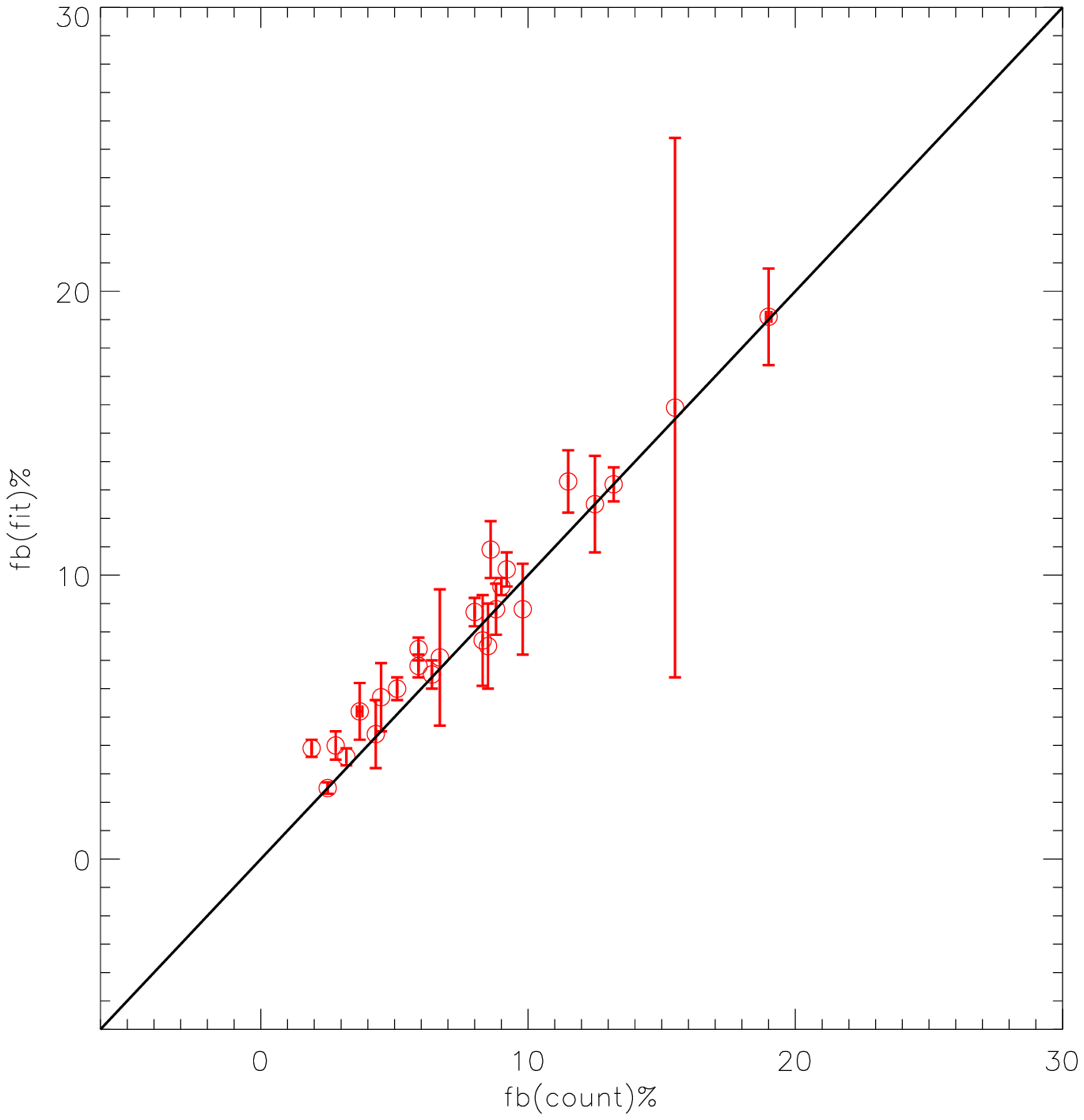}
\caption{Comparisons of binary fractions from the whole field of view between different methods. Blue filled circles: binary fractions obtained by high q method are compared to the counting method. Red open circles: binary fractions obtained by fitting method are compared to the counting method. The black solid line shows where the two methods give the same results.}
\label{fig:compAll}
\end{figure}

From the table,  we can calculate the mean binary fractions including 25 reliable binary fractions from the three methods: 5.2\% (high q), 6.3\% (counting), and 7.3\% (fitting), respectively. In Figure ~\ref{fig:compAll}, we show the comparisons of binary fractions obtained through different methods. 
Blue filled circles are for binary fractions obtained by high q method compared to the counting method. Red open circles are for binary fractions obtained by the fitting method compared to the counting method. The black solid line shows where the two methods give the same results. 
From the figure, we can see that the binary fractions obtained through the counting and fitting method are consistent with each other, while they are usually larger than the values obtained through the high mass-ratio method. This is because the latter method does not include binaries with small mass ratio that are closer to the main sequence, while the former methods can statistically recover part of the signals from the small mass-ratio binaries.

For the fitting method in Table ~\ref{tab:whole}, the fitted power $x$ has a mean value of $-1.9\pm1.1$ (from the 25 reliable clusters), suggesting that most binaries tend to have small mass ratios. The fitted minimum binary fraction $q_{min}$ has a mean value of $0.30\pm0.06$, which is smaller than the cut-off ratio in the high mass-ratio method ($q=0.5$), indicating that the fitting method can recover part of mass-ratio binaries that are hidden in the main sequence.

The comparison between the minimum whole-field binary fractions with other efforts 
is not as good as one would expect, given that they are using the same data sets.
The most direct comparison uses the binary fraction above a certain value of $q$, such as
0.5 for \citet{milone12} and this work, which was chosen because the distance of these binaries from
the Main Sequence ridge line is generally above the 3$\sigma$ uncertainties of a star.
\citet{sollima07} also presents
a binary fraction above some minimum binary ration, their $q_{min}$, but they do not give
values for $q_{min}$, which may differ between clusters (their Table 3).

For the 13 globular clusters considered by Sollima, Terzan 7 and Palomar 12 have binary fractions
that are significantly higher than the other clusters, a result confirmed by Milone and this work,
although the values differ between the groups.  For the remaining 11 clusters, we do not find a 
meaningful relationship between the binary fractions from Sollima and Milone (Figure 2).  The values from
Sollima are systematically larger than those of Milone, but that may be due to a difference
in how their binary fractions are calculated.  There is the same offset and lack of a correlation
when comparing the Sollima values to our values (Figure 2).  
When comparing our values with those of Milone,
there is no systematic offset and most of the values are consistent with each other, but there
are a few significant discrepancies, which we discuss further below (Figure 3).

Although these various binary 
fractions are determined from the same data sets, the number of stars used is somewhat different
and the photometric software is different.  However, each of the photometric methods has a 
heritage and is well-tested, although the method used by Milone \citep{anderson08} appears to
be more efficient and more magnitudes are extracted for the same fields in approximately the
same magnitude ranges.  One might expect that the different photometric methods would lead to
different numbers of stars extracted but would not lead to systematically incorrect magnitudes on 
a scale greater than 3$\sigma$.  Furthermore, the various groups performed simulations to
demonstrate the reliability of recovering artificial stars that were placed into the very fields
being analyzed.  For most of these clusters, the other corrections are too small to account for 
the differences, such as overlapping stars and field star contamination.
The reasons for these differences remain unclear and a cross-method investigation is beyond the
scope of this effort.

\subsection{Binary Fractions within the Half-mass Radius}

The analysis for the binary fractions within the half-mass radius has physical importance in that it provides a uniform basis to make comparisons between clusters and with theoretical models.  One difference with the fitting method is that we fixed the values of the power index $x$ and the minimum mass-ratio $q_{min}$ to the ones obtained by whole field analysis, because there are more stars to constrain the parameters using the whole chips.

The binary fractions are listed in Table ~\ref{tab:rh}, which is similar to Table ~\ref{tab:whole}, except that we omit the size column. We also exclude the columns of power index $x$ and the minimum mass-ratio $q_{min}$, as they are all fixed to the values in Table ~\ref{tab:whole}. The average binary fractions within the half-mass radius from this table is 5.6\% (for the $q>0.5$ binary fractions), which includes 27 reliable clusters with flags of d, n, or g.  

There is a very tight relationship between the fitting and the counting method, so either are equivalently good for further analysis).  However, there is a poorer relationship between the high q binary fraction and the one from the counting method (or from the fitting method).  This is also a poorer relationship than when using the whole fields.  Based on the tests
conducted in Paper I, the weaker relationship cannot be attributed to uncertainties in the
methods unless there are mitigating factors, such as an incorrect estimate for the foreground/background stellar contamination, significant differential reddening, or multiple populations.
If these mitigating factors can be ignored, we would attribute the binary fraction difference to an increasing binary fraction in the range $0.3 \leq q \leq 0.5$, which are included in the counting method but not in the high-q method.

 
From the binary fractions that we calculated, we examine whether they are correlated with the physical properties of age, dynamical age (age divided by the relaxation time at r$_h$), metallicity ([$Fe/H$]), and absolute magnitude M$_V$.  To make the comparison, we assume that the binary fraction and above quantities are linearly related, as
\begin{eqnarray}
y = b+k*x,
\end{eqnarray}
where y is the binary fraction and x is the different properties of globular clusters.

Figure ~\ref{fig:rh_Fb} shows the relationships between the half-mass binary fractions and the ages, dynamical ages (absolute ages divided by the relaxation time at half-mass radius), metallicity [Fe/H], and the absolute V magnitudes. The fitted line parameters and the non-parametric Spearman rank coefficients are shown in Table ~\ref{tab:fb_rh}. 

The strongest correlation is between the half-mass binary fractions and the cluster absolute V magnitudes, meaning that less luminous (massive) clusters have higher binary fractions. This correlation is also confirmed by the Spearman rank coefficient, with reliable significance. This relationship was also discovered by \citet{milone08, milone12}.
The lowest luminosity globular cluster, E3, is not included in our work, but as evident in \citet{milone12}, it supports the increase of binary fraction with decreasing luminosity.
The lower density end of this trend would correspond to the open clusters, where they usually have higher binary frequencies ($\sim 30\%$ for Hyades by \citet{griffin88}, $>59$ \% for Praesepe by \citet{bolte91}, and 50\% for M67 by \citet{fan96}). A more recent study of 9 open clusters show 15\%-54\% of binary fraction \citep{bica05}. For field stars near the solar neighborhood, the binary frequency can be as high as 50\% \citep{Duquennoy91,Halbwachs03}.

From this table, we can see that the half-mass binary fractions appears to be anti-correlated with the cluster ages at above the 95\% confidence level.  However, two clusters have a significant effect on this result, Terzan 7 and Palomar 12, which are both relatively young and have low absolute luminosities.  
We return to this relationship when considering the core binary fractions, below.
We do not see a correlation of the half-mass radius binary fractions with either the dynamical time or the metallicity [Fe/H], which is the same result as that of \citet{milone12}, who used a larger sample but did the analysis with binary fractions within the core radius as well as between the core and the half light radius.

\section{Binary Fractions within the Core Regions}

The binary fractions within the core regions are expected to respond most rapidly to dynamical 
interactions between stars.  In our sample, not all the clusters are suitable for the core binary fraction analysis, because the core regions for some dense clusters are too crowded to recover stars
or the core regions for some core-collapsed clusters are too small to include enough stars from
which to measure a binary fraction. 
With such limitations (and selection effects), we analyzed 25 clusters in our sample with the high mass-ratio method.

Table ~\ref{tab:coreFb} shows the core binary fractions for 25 clusters with the high mass-ratio ($q>0.5$) method. The mean binary fractions within the core regions is about 7.0\% (excluding 8 clusters with field star contamination or due to large photometric errors and low star numbers). 
The correlation of these binary fractions with those of Milone is a good but not 
perfect correlation, with a few clusters in which our binary fractions are sytematically higher.
The relationship between the binary fraction and the physical properties of the clusters is 
similar to the results using the binary fractions within r$_h$, with one exception.  
There is a strengthening of the inverse correlation between the binary fraction and the cluster 
age (Figure ~\ref{fig:coreFb}), and this relationship remains significant even when the two youngest clusters are excluded.
Our ages $t_{Gyr}$, as reported in Paper I, are mainly from \citet{salaris02}, although a few additional
ages were used from \citet{fullton96}, \citet{alonso-Garcia11}, and \citet{sollima07}.
The correlation coefficient of the correlation is -0.595, a significance of 99.8\%, and with the 
values $k = -2.40 \pm 0.14$ and $b = 32.8 \pm 1.6$.
We can examine this relationship using a more uniform set of ages, where \citet{marin09} determined the absolute magnitudes of the main sequence turnoff, and using isochrones, obtained a relative age scale.  The relative age scale is normalized at the mean, so one can multiply it by the appropriate age (e.g., 12 Gyr) to obtain an absolute age.
The results are not strongly dependent on the isochrone library, so we use their D07$_{ZW}$ values and find a similar anticorrelation (Figure ~\ref{fig:coreFb2}) with slope of $-30.3 \pm 4.8$ (for a mean age of 12 Gyr, this would correspond to $k = -2.52 \pm 0.40$) and intercept $b = 34.9 \pm 4.6$. 
The Spearman correlation indicates a significance at the 99.4\% level, although if the points are normally distributed, we could use the Pearson correlation test, which yields a higher levels of significance, $5.3 \sigma$.
In this case, we included clusters for which the formal binary fraction is negative (mainly systems with large numbers of foreground/background stars), except NGC 6656, where the field stars are so dominant that the binary fraction is unreliable.  However, the result does not change significantly if those points are excluded (or if NGC 6656 were included).
This result has implications that we comment upon below.

\section{The Radial Distribution of the Binary Fractions}\label{sec:fb_r}

Another important prediction of the models is that there will be a significant
radial dependence on the binary fraction as well as on the stellar mass function.
This was explored by both Sollima and Milone, who both found radial gradients
in the binary fraction.  We study the radial distribution of the binary fraction and also find evidence for a radial gradient. In addition we investigate whether this is an intrinsic property or one that developed due to the dynamical interactions in the cluster.

In deriving the radial distribution of the binary fractions, we divide the cluster
up into three radial bins of approximately equal numbers of stars, so that the 
uncertainties are comparable in the three bins.  The uncertainties are greater 
when compared to the whole field values, due to the fewer number of stars, and 
for the inner bin in clusters with dense central regions, the results have greater uncertainty. 
The methods that involve fitting, $f_b(count)$ and $f_b(fit)$ have larger uncertainties
than $f_b(q >  0.5)$ due to the additional components that must be constrained.
The results are given in Table 4.  Due to the smaller uncertainties,
we use $f_b(q >  0.5)$ to examine the radial binary dependence

In Figure ~\ref{fig:Fb_r}, we plot all the high mass-ratio ($q>0.5$) binary fractions 
at different annular bins from all the clusters in our sample as a function of their 
radial distances from the cluster center. 
The left panel is the high mass-ratio binary fractions normalized to their core fractions,
and the distances are normalized to their core radii. 
There are only 17 clusters with reliable core binary fractions measured in our sample. 
The straight line fitted results are shown in Table ~\ref{tab:Fb_r}, first row, which 
shows a moderate correlation (slope $k=-0.41$). 
The Spearman rank correlation coefficient, however, shows a significant correlation 
(coefficient of $=-0.69$, with a highly reliable significance).

The right panel in Figure ~\ref{fig:Fb_r} and the second row in Table ~\ref{tab:Fb_r} 
show the high mass-ratio binary fractions normalized to their half-mass fractions 
as a function of the distances normalized to their half-mass radii. 
The straight line fit shows a strong correlation (slope $k = -0.99$), with a Spearman  
rank correlation coefficient of -0.61, designating a highly significant correlation.  
Even though there can be significant differences between investigators for the value of a binary
fraction for the same cluster, the radial distributions are very similar between investigators.

When a result is found, such as the radial decrease in the binary fraction, it could be due
to evolutionary forces, but it could also be due to initial conditions -- the cluster was simply
born with that property.  
To examine this issue, we compare the six clusters where the ratio of the age to dynamical
relaxation time (at $r_h$) is less than four, NGC 104, NGC 5053, NGC 5272 (M3), NGC 5466, NGC 5897, and Arp 2.  If the radial decrease steepens with the number of elapsed relaxation times, these six clusters should show shallower radial binary fraction slopes.
Of these six, only two show a radial decrease where the binary fraction drops by at least 20\%.
For the remaining 25 clusters for which we have good radial information, only four fail to
show a radial decrease of 20\% or greater.
Clusters with dynamical ages above four relaxation times show a radial decrease in 84\% of the
cases while those with younger ages show the decrease 33\% of the time.
This implies that the decrease in the binary fraction with radius is due to evolution rather than birth.

\section{Discussion and Conclusions}

The astrophysical question we sought to address is whether globular clusters follow the 
predictions of the dynamical models.  As a cluster progresses through several relaxation times,
mass segregation should occur and soft binaries are rendered unbound.  Mass segregation 
leads to binaries sinking deeper into the cluster, increasing the binary fraction, but 
strong dynamical interactions can destroy binaries, having the opposite effect.  
To understand which effect prevails requires both theoretical and observational efforts,
with the theoretical work indicating that the core binary fraction will rise with time prior 
to the core collapse period (Fregau 2009).

Another consideration in comparing theory to observation is that the initial properties of
a cluster can mimic that expected from dynamical evolution.  For example, Frank (2013) 
studied the globular cluster Palomar 14, which has a long half-mass relaxation time (20 Gyr) 
and lies in the outer-halo, where it does not feel strong tidal forces from the Galaxy.
They find that Palomar 14 exhibits significant mass segregation despite having an age less than 
even a single relaxation time.  The most likely conclusion is that it was born as a mass-
segregated cluster, and if this formation history is common, it complicates the ability to
compare data to models.  A final consideration is that the cluster-to-cluster dispersion may
be larger than the effect one is trying to measure, so it is helpful to understand the
range of initial conditions in a globular cluster.

A significant result is that the core binary fraction decreases inversely with 
the age of the globular clusters.  This was first found by \citet{sollima07} also at a 
confidence level exceeding 99\%; both we and Sollima used ages from \citet{salaris02}.  
However, when we used ages from \citet{marin09}, we obtain the same result, with only a slight decrease in significance.
When \citet{milone12} used the ages from \citet{marin09}, or from \citet{salaris02}, they did not find a significant anticorrelation when using their entire sample.  
Restricted to just lower density clusters, they do not find an anticorrelation when using the relative ages from \citet{marin09}, but they appear to find an anticorrelation (their Figure B,10) when using the absolute ages from \citet{salaris02} and \citet{deang2005}.
Since the anticorrelation is with age and not dynamical age, it suggests that the 
younger clusters were born with a higher binary fraction.  
Evidently, the conditions from which younger globular clusters formed were more conducive to binary formation.

A examination of the binary fraction as a function of dynamical time can be used to obtain insight
into the likely range of initial conditions.
To examine this issue, we use the core binary fraction from Milone for 
$q > 0.5$ in Figure ~\ref{fig:binary_dyn_t}, which leads to a more statistically significant
result than using the binary fractions presented in this paper. 
At the shortest dynamical times, $t/t_{r,h} < 3$, the range in the binary fraction is
about an order of magnitude.  This is probably representative of the range of initial binary 
fractions in clusters, although the absolute value of the binary fraction must have been larger for a few reasons.  
Binaries where one or both stars are remnants (white dwarfs or neutron stars) are recorded as single
stars in the CMD method, while many of the soft binaries are destroyed even by the first 
relaxation time.

Another impression one has from this figure is that the binary fraction lies within 
an envelope that is slowly decreasing with dynamical age, by about a factor of two over the 
range of the figure.  That is, above 10 relaxation times,
there are no clusters with binary fractions above 6\% and nearly half have binary fractions
below 2\%.  This is in contrast to the earlier times, where larger binary fractions are 
relatively more common.  It will be difficult to improve on this data set as the HST images
are of excellent quality and the best globular clusters have been observed.

The radial decrease in the binary fraction is a robust result and one can account for an
intrinsic dispersion in the binary fraction by dividing by the binary fraction either in the
core or within half light radius.  This radial decrease is nearly always seen in clusters 
older than four dynamical times and seen less commonly for shorter dynamical times.
This implies that the radial binary fraction distribution is a result of relaxation and
tidal effects rather than due to initial conditions.

The mass function for the binaries is still not very well known.  For $q > 0.5$, one can simply
count the stars in the color-magnitude diagram to obtain a binary fraction, but to
extend this to lower mass ratios, a parametric fitting method is needed because the magnitude
of the departure from the Main Sequence from photometry errors becomes comparable to that
due to the pairing of two stars.  
Our fitting procedure finds that the number of binaries increases with decreasing mass ratio
(down to $q = 0.3$), although there is significant uncertainty in the value of the exponent.
This would be consistent with models in which binary stars are random pairs of stars, yet the
work by Milone indicates that the mass function is flat above $q = 0.5$.
We caution the reader that our fitting procedure depends upon an assumed Gaussian symmetry 
in the width of the main sequence.  If this assumption is not generally valid, then our result
of a rising binary mass distribution is thrown into question.

There are other avenues for studying binary populations in star clusters and one of the most
effective is through spectroscopic programs.  A number of these clusters have stars bright 
enough for ground-based studies on large telescopes with multi-object capabilities.  
Sufficient spectroscopic monitoring can determine the period and velocity amplitude of the 
binary star system, which determines the mass function.  With the accurate photometric measurements
from the HST data sets and a mapping between the absolute magnitude and stellar mass, the
orbital parameters can be deduced.  Such data provide information on the hardness of the binaries,
essential data not provided by the CMD method used here.  

Spectroscopy will help in another important area in which a star has a degenerate partner, which
will lead to a periodic shift in the stellar absorption lines.  This will help to give a fuller
picture for the total number of stars in binaries, and these are probably binaries that formed
relatively early.  To help to facilitate such spectroscopic
programs, we compiled a list of 6421 binaries with $q > 0.5$, with RA, DEC, magnitudes, and 
distance from the center of the globular cluster (Ji 2011; also available from the authors upon request).
The photometry for all stars in the fields is available from the ACS Globular Cluster 
Treasury Program, with links from the MAST website (their magnitudes are slightly different from ours).

\section{Acknowledgements}
The authors would like to thank A.E. Dolphin for answering our many questions that arose when using the photometry package Dolphot. 
For their many questions and suggestions, we would like to thank from Mario Mateo, Jon Miller, Eric Bell, Sally Oey, Patrick Seitzer, and a very helpful referee.  We gratefully acknowledge financial support through HST programs 10939 and 11125 from NASA. 


{}

\clearpage

\begin{deluxetable}{ccrrrrcrc}
\rotate
\tablecolumns{9}
\tablewidth{0pc}
\tablecaption{Fitting results for whole field of view}
\tablehead{
Source & size(in $r_h$) & $f_b(q>0.5)\%$ & $f_b(count)\%$ & $f_b(fit)\%$ & Power x & $q_{min}$ & $\chi^2$/dof &Note$^a$ }
\startdata
    NGC104 &  0.75 &    3.01$\pm$0.13 &    8.04$\pm$0.70 &    8.70$\pm$0.50 &   -2.95$\pm$0.05 &    0.28$\pm$0.01  &  204.1/162 &      d\\
    NGC288 &  1.06 &    6.47$\pm$0.31 &   12.87$\pm$1.32 &   13.20$\pm$0.60 &   -1.50$\pm$0.09 &    0.27$\pm$0.00  &  120.1/106 &      g\\
    NGC362 &  2.89 &    4.39$\pm$0.16 &    8.28$\pm$0.75 &    9.60$\pm$0.30 &   -2.11$\pm$0.19 &    0.24$\pm$0.01  &  176.9/142 &      d\\
   NGC1851 &  4.64 &    2.88$\pm$0.15 &    5.51$\pm$0.82 &    6.50$\pm$0.50 &   -2.91$\pm$0.05 &    0.29$\pm$0.01  &  126.4/136 &      d\\
   NGC2808 &  2.96 &    1.26$\pm$0.18 &   -2.05$\pm$0.66 &    0.50$\pm$0.00 &    0.00$\pm$3.00 &    0.77$\pm$0.45  &  299.5/164 &    d,p,e\\
   NGC4590 &  1.57 &    8.12$\pm$0.30 &   11.14$\pm$1.27 &   13.30$\pm$1.10 &   -2.48$\pm$0.38 &    0.33$\pm$0.00  &  141.6/ 96 &      g\\
   NGC5053 &  0.91 &    5.57$\pm$0.40 &    8.09$\pm$1.81 &    7.50$\pm$1.50 &   -0.98$\pm$0.75 &    0.33$\pm$0.01  &   43.9/ 70 &      g\\
        M3 &  1.02 &    5.10$\pm$0.17 &    5.45$\pm$0.70 &    6.80$\pm$0.40 &   -2.11$\pm$0.19 &    0.33$\pm$0.00  &  200.3/161 &      d\\
   NGC5466 &  1.03 &    5.19$\pm$0.35 &    8.96$\pm$1.86 &    8.80$\pm$1.60 &   -1.50$\pm$0.38 &    0.30$\pm$0.03  &   34.6/ 63 &      g\\
   NGC5897 &  1.15 &    5.74$\pm$0.28 &    6.10$\pm$1.27 &    7.10$\pm$2.40 &   -0.75$\pm$0.19 &    0.33$\pm$0.01  &  100.4/ 98 &      g\\
   NGC5904 &  1.34 &    3.01$\pm$0.14 &    4.61$\pm$0.82 &    5.70$\pm$1.20 &   -3.00$\pm$0.05 &    0.30$\pm$0.06  &  113.6/122 &      d\\
   NGC5927 &  2.15 &    2.44$\pm$0.21 &   12.29$\pm$0.93 &   14.20$\pm$0.50 &   -3.00$\pm$0.01 &    0.24$\pm$0.01  &  157.3/148 &      f\\
   NGC6093 &  3.88 &    3.87$\pm$0.18 &    1.42$\pm$0.84 &    3.90$\pm$0.30 &   -3.00$\pm$0.09 &    0.40$\pm$0.00  &  207.6/125 &      d\\
   NGC6121 &  0.55 &    4.78$\pm$0.48 &   11.73$\pm$2.16 &   16.20$\pm$0.90 &   -3.00$\pm$0.19 &    0.25$\pm$0.01  &   66.0/ 62 &      f\\
   NGC6101 &  2.25 &    5.33$\pm$0.23 &    6.54$\pm$1.12 &    7.40$\pm$0.40 &   -1.50$\pm$0.19 &    0.31$\pm$0.00  &  105.8/100 &      g\\
       M13 &  1.40 &    3.28$\pm$0.14 &    2.15$\pm$0.71 &    2.50$\pm$0.20 &   -2.81$\pm$0.09 &    0.41$\pm$0.01  &  147.1/145 &      d\\
   NGC6218 &  1.34 &    3.15$\pm$0.22 &    8.51$\pm$1.29 &    8.80$\pm$0.90 &   -1.12$\pm$0.38 &    0.18$\pm$0.01  &   92.6/ 91 &      g\\
   NGC6341 &  2.32 &    4.12$\pm$0.18 &    4.19$\pm$0.69 &    4.40$\pm$1.20 &   -1.03$\pm$0.75 &    0.27$\pm$0.03  &  247.0/146 &      d\\
   NGC6352 &  1.15 &   -0.58$\pm$0.87 &    6.92$\pm$2.09 &    5.50$\pm$2.30 &   -1.88$\pm$0.38 &    0.21$\pm$0.06  &  110.4/ 99 &      f\\
   NGC6362 &  1.15 &    4.39$\pm$0.29 &   11.21$\pm$1.42 &   12.50$\pm$1.70 &   -2.48$\pm$0.09 &    0.24$\pm$0.03  &  108.2/ 95 &      g\\
   NGC6397 &  0.82 &    4.48$\pm$0.49 &    7.81$\pm$1.81 &   10.90$\pm$1.00 &   -3.00$\pm$0.19 &    0.35$\pm$0.00  &  103.1/ 79 &      g\\
   NGC6541 &  2.23 &    2.53$\pm$0.19 &    5.57$\pm$0.78 &    7.60$\pm$0.30 &   -3.00$\pm$0.02 &    0.27$\pm$0.00  &  227.7/140 &     d,f\\
   NGC6624 &  2.89 &    1.57$\pm$0.54 &   13.77$\pm$1.32 &   27.30$\pm$2.90 &   -2.98$\pm$0.19 &    0.18$\pm$0.03  &  200.2/144 &     d,f\\
   NGC6637 &  2.82 &    2.06$\pm$0.30 &    5.53$\pm$1.10 &    6.60$\pm$0.40 &   -1.88$\pm$0.19 &    0.18$\pm$0.01  &  142.3/135 &     d,f\\
   NGC6652 &  4.93 &    0.87$\pm$0.71 &   -0.03$\pm$2.08 &    1.50$\pm$0.50 &    0.00$\pm$3.00 &    0.89$\pm$0.45  &   92.7/ 85 &     d,f\\
   NGC6656 &  0.70 &   -4.87$\pm$0.28 &   -0.69$\pm$0.97 &    3.90$\pm$1.40 &   -3.00$\pm$0.05 &    0.18$\pm$0.03  &  488.0/124 &      f\\
   NGC6723 &  1.55 &    4.55$\pm$0.19 &    8.79$\pm$0.97 &   10.20$\pm$0.60 &   -3.00$\pm$0.05 &    0.28$\pm$0.00  &  113.2/118 &      g\\
   NGC6752 &  1.24 &    0.91$\pm$0.16 &    2.97$\pm$1.04 &    4.00$\pm$0.50 &   -2.81$\pm$0.19 &    0.25$\pm$0.01  &  124.2/ 98 &      g\\
   Terzan7 &  3.07 &   12.23$\pm$0.79 &   19.67$\pm$2.62 &   19.10$\pm$1.70 &    0.00$\pm$0.19 &    0.18$\pm$0.06  &   40.5/ 63 &      g\\
      Arp2 &  1.34 &    8.51$\pm$0.55 &    8.32$\pm$2.16 &    7.70$\pm$1.60 &    0.00$\pm$0.38 &    0.44$\pm$0.23  &   57.9/ 74 &      g\\
   NGC6809 &  0.84 &    3.31$\pm$0.22 &    3.10$\pm$1.31 &    3.60$\pm$0.30 &    0.00$\pm$0.75 &    0.25$\pm$0.06  &   64.8/ 77 &      g\\
   NGC6981 &  2.54 &    5.33$\pm$0.24 &    5.50$\pm$1.25 &    6.00$\pm$0.40 &   -1.88$\pm$0.75 &    0.38$\pm$0.01  &   85.3/102 &      d\\
   NGC7078 &  2.37 &    4.38$\pm$0.17 &    0.50$\pm$0.62 &    1.50$\pm$0.10 &    0.00$\pm$0.19 &    0.25$\pm$0.03  &  229.2/160 &     d,e\\
   NGC7099 &  2.30 &    3.06$\pm$0.28 &    4.23$\pm$1.08 &    5.20$\pm$1.00 &   -3.00$\pm$0.75 &    0.33$\pm$0.06  &   85.7/ 68 &      g\\
 Palomar12 &  1.38 &   13.44$\pm$1.26 &    7.99$\pm$4.25 &   15.90$\pm$9.50 &   -0.75$\pm$1.50 &    0.25$\pm$0.45  &   13.1/ 33 &      n\\

\enddata

\tablecomments{
Note The results quality flags are d: dense core; f: field stars; n: small number of stars; p: multi-populations; e: large photometric errors; g: good estimate. Binary fractions with flags of g, d, and n are usually good to use. Uncertainty from flag f is quite large, so be caution when use those values. Binary fractions with more than one flag are not good to use.}

\label{tab:whole}
\end{deluxetable}
\clearpage

\begin{deluxetable}{crrrcc}
\tablecolumns{6}
\tablewidth{0pc}
\tablecaption{Fitting results within half light radius}
\tablehead{
Source & $f_b(q>0.5)\%$ & $f_b(count)\%$ & $f_b(fit)\%$ & $\chi^2$/dof& Note$^a$
}
\startdata
    ngc104 &    3.03$\pm$ 0.13 &    8.03$\pm$0.70 &    8.73$\pm$0.81 &   205.6/162 &     d\\
    ngc288 &    6.29$\pm$ 0.32 &   12.26$\pm$1.38 &   11.85$\pm$0.36 &    96.9/ 99 &     g\\
    ngc362 &    5.69$\pm$ 0.36 &    8.17$\pm$1.30 &    6.62$\pm$0.35 &   122.8/118 &     d\\
   ngc1851 &    3.89$\pm$ 0.85 &   -1.06$\pm$2.79 &    0.50$\pm$0.00 &    40.3/ 73 &    d,e\\
   ngc2808 &  -12.09$\pm$ 0.96 &   -4.84$\pm$2.44 &    0.50$\pm$0.00 &    71.9/ 90 &   d,e,p\\
   ngc4590 &    6.24$\pm$ 0.30 &    7.33$\pm$1.39 &    8.62$\pm$1.21 &   112.3/ 87 &     g\\
   ngc5053 &    5.57$\pm$ 0.40 &    8.17$\pm$1.81 &    8.25$\pm$1.68 &    42.5/ 70 &     g\\
        m3 &    5.03$\pm$ 0.17 &    5.42$\pm$0.70 &    6.62$\pm$0.24 &   197.4/161 &     d\\
   ngc5466 &    5.23$\pm$ 0.35 &    9.03$\pm$1.86 &    9.36$\pm$0.63 &    33.4/ 63 &     g\\
   ngc5897 &    5.63$\pm$ 0.28 &    5.84$\pm$1.28 &    6.28$\pm$0.30 &    82.5/ 96 &     g\\
   ngc5904 &    3.15$\pm$ 0.15 &    4.84$\pm$0.85 &    5.48$\pm$1.17 &   111.6/122 &     d\\
   ngc5927 &    3.87$\pm$ 0.28 &    6.18$\pm$1.29 &    5.18$\pm$1.03 &   107.8/127 &     f\\
   ngc6093 &    7.59$\pm$ 0.58 &    1.25$\pm$1.78 &    2.22$\pm$1.73 &    80.8/ 98 &     d\\
   ngc6121 &    4.79$\pm$ 0.48 &   -1.94$\pm$3.37 &    0.50$\pm$0.35 &    46.7/ 40 &     g\\
   ngc6101 &    5.50$\pm$ 0.38 &    6.59$\pm$1.12 &    7.70$\pm$0.65 &   105.7/100 &     g\\
       m13 &    3.46$\pm$ 0.16 &    2.12$\pm$0.75 &    2.56$\pm$0.60 &   149.2/144 &     d\\
   ngc6218 &    3.21$\pm$ 0.24 &    8.31$\pm$1.35 &    8.87$\pm$2.94 &    93.3/ 88 &     g\\
   ngc6341 &    5.31$\pm$ 0.29 &    3.27$\pm$0.98 &    3.33$\pm$0.97 &   151.0/136 &     d\\
   ngc6352 &   -0.57$\pm$ 0.88 &    6.06$\pm$2.11 &    5.91$\pm$1.53 &   104.7/ 98 &     f\\
   ngc6362 &    4.31$\pm$ 0.29 &   10.64$\pm$1.44 &   11.59$\pm$1.04 &    99.9/ 94 &     g\\
   ngc6397 &    4.47$\pm$ 0.49 &    7.75$\pm$1.81 &   10.84$\pm$0.67 &   100.7/ 79 &     g\\
   ngc6541 &    3.94$\pm$ 0.28 &    1.69$\pm$1.08 &    2.96$\pm$0.40 &   184.6/123 &     d\\
   ngc6624 &    1.31$\pm$ 0.57 &    8.13$\pm$1.96 &    5.42$\pm$7.96 &    63.7/ 93 &     f,d\\
   ngc6637 &    4.16$\pm$ 0.40 &    4.10$\pm$1.63 &    2.96$\pm$0.59 &   102.5/103 &     f,d\\
   ngc6652 &    3.17$\pm$ 0.78 &   11.82$\pm$3.67 &    3.45$\pm$1.70 &    42.6/ 41 &     f,d\\
   ngc6656 &   -4.94$\pm$ 0.28 &   -0.69$\pm$0.97 &    3.76$\pm$1.39 &   486.9/124 &     f\\
   ngc6723 &    4.66$\pm$ 0.21 &    7.19$\pm$1.06 &    8.37$\pm$1.03 &   101.3/115 &     g\\
   ngc6752 &    0.74$\pm$ 0.16 &    2.26$\pm$1.07 &    2.96$\pm$0.45 &   118.3/ 96 &     g\\
   Terzan7 &   17.15$\pm$ 1.40 &   13.56$\pm$2.71 &   13.30$\pm$2.03 &    43.9/ 62 &     g\\
      Arp2 &    8.05$\pm$ 0.55 &    4.92$\pm$3.31 &    4.44$\pm$1.39 &    30.9/ 50 &     g\\
   ngc6809 &    3.31$\pm$ 0.22 &   -5.14$\pm$2.92 &    0.50$\pm$0.06 &    33.4/ 43 &     g\\
   ngc6981 &    7.18$\pm$ 0.38 &    5.43$\pm$1.27 &    6.13$\pm$0.64 &    85.7/102 &     d\\
   ngc7078 &    5.85$\pm$ 0.32 &    0.47$\pm$0.62 &    1.48$\pm$0.17 &   231.0/160 &     d\\
   ngc7099 &    3.78$\pm$ 0.44 &    2.50$\pm$1.52 &    0.50$\pm$1.25 &   103.6/ 62 &     d\\
 Palomar12 &   13.36$\pm$ 1.30 &   -8.38$\pm$6.02 &   12.31$\pm$19.88 &    7.2/20.0 &     g\\
\enddata
\tablecomments{Result flags as in Table ~\ref{tab:whole}.  }
\label{tab:rh}
\end{deluxetable}
\clearpage

\begin{deluxetable}{cccccc}
\tablecolumns{6}
\tablewidth{0pc}
\tablecaption{Fitting results for the half-mass binary fractions Vs. different properties of clusters}
\tablehead{
Vs. Properties &  k & b & $\chi^2/dof$ & coefficient$_r$ & significance
}
\startdata
Age &-0.53$\pm$0.05&10.02$\pm$0.61&   1055.2/ 25&-0.392&0.043\\
Dynamical Age &-0.10$\pm$0.01& 4.65$\pm$0.09&   1042.0/ 25&-0.068&0.737\\
$[Fe/H]$&-1.04$\pm$0.11& 2.44$\pm$0.18&   1054.4/ 25&-0.016&0.936\\
Mv & 0.63$\pm$0.06& 9.12$\pm$0.41&   1017.0/ 25& 0.490&0.010\\
\enddata
\label{tab:fb_rh}
\end{deluxetable}


\begin{deluxetable}{crrrrrc}
\tablecolumns{7}
\tablewidth{0pc}
\tablecaption{Fitting results for radial bins}
\tablehead{
Source & Bin Range$(r_h)$ & $f_b(q>0.5)\%$ & $f_b(count)\%$ & $f_b(fit)\%$ &  $\chi^2$/dof &Note$^a$
}
\startdata
    ngc104 &  0.00-0.29 &    4.14$\pm$0.29 &    5.63$\pm$1.27 &    4.19$\pm$0.38 &   144.7/130 &     d\\
 &  0.29-0.42 &    2.72$\pm$0.22 &   10.02$\pm$1.25 &   11.94$\pm$0.54 &    75.7/110 &     d\\
 &  0.42-0.56 &    2.11$\pm$0.18 &    6.22$\pm$1.26 &   10.59$\pm$1.63 &   104.7/100 &     d\\
    ngc288 &  0.00-0.43 &    4.75$\pm$0.51 &    4.51$\pm$2.39 &    2.59$\pm$0.98 &    64.0/ 63 &     g\\
 &  0.43-0.70 &    6.96$\pm$0.58 &   17.21$\pm$2.33 &   23.63$\pm$2.32 &    77.0/ 71 &     g\\
 &  0.70-1.05 &    6.94$\pm$0.55 &   18.69$\pm$2.32 &   35.45$\pm$8.90 &    38.8/ 67 &     g\\
    ngc362 &  0.00-0.97 &    5.74$\pm$0.37 &    8.38$\pm$1.34 &    6.53$\pm$0.40 &   127.7/118 &     d\\
 &  0.97-1.48 &    4.57$\pm$0.27 &   12.61$\pm$1.32 &   15.76$\pm$1.15 &    91.9/101 &     d\\
 &  1.48-2.16 &    2.53$\pm$0.22 &    7.73$\pm$1.33 &   10.37$\pm$1.14 &    89.0/ 87 &     g\\
   ngc1851 &  0.00-1.55 &    3.80$\pm$0.37 &    3.03$\pm$1.47 &    2.84$\pm$0.63 &    91.5/113 &     d\\
 &  1.55-2.34 &    2.20$\pm$0.25 &    5.93$\pm$1.45 &    8.62$\pm$1.02 &    76.0/ 87 &     d\\
 &  2.34-3.55 &    1.50$\pm$0.20 &    6.48$\pm$1.45 &    8.74$\pm$0.97 &    79.1/ 76 &     d\\
   ngc2808 &  0.00-1.35 &   -1.56$\pm$0.38 &   -2.44$\pm$1.19 &    0.50$\pm$0.01 &   186.9/141 &   d,e,p\\
 &  1.35-1.73 &    1.14$\pm$0.29 &   -1.08$\pm$1.17 &    0.56$\pm$0.14 &   180.3/132 &   d,e,p\\
 &  1.73-2.26 &    1.49$\pm$0.28 &   -3.54$\pm$1.17 &    0.50$\pm$0.02 &   204.4/130 &   d,e,p\\
   ngc4590 &  0.00-0.44 &    8.27$\pm$0.61 &    7.02$\pm$2.30 &    5.42$\pm$1.21 &    47.3/ 64 &     g\\
 &  0.44-0.74 &    3.88$\pm$0.45 &    5.21$\pm$2.30 &    6.90$\pm$1.58 &    38.4/ 55 &     g\\
 &  0.74-1.14 &    6.40$\pm$0.51 &   10.83$\pm$2.28 &   19.70$\pm$1.83 &    74.5/ 60 &     g\\
   ngc5053 &  0.00-0.36 &    5.64$\pm$0.77 &    8.88$\pm$3.24 &    5.91$\pm$2.57 &    22.2/ 43 &     g\\
 &  0.36-0.52 &    4.71$\pm$0.70 &    2.33$\pm$3.31 &    6.41$\pm$4.18 &    26.6/ 40 &     g\\
 &  0.52-0.68 &    4.90$\pm$0.74 &    0.12$\pm$3.32 &    5.42$\pm$2.20 &    25.1/ 41 &     g\\
        m3 &  0.00-0.35 &    6.17$\pm$0.39 &    6.48$\pm$1.26 &    5.44$\pm$1.48 &   112.2/135 &     d\\
 &  0.35-0.54 &    3.94$\pm$0.28 &    6.85$\pm$1.25 &    9.21$\pm$0.38 &    82.8/111 &     d\\
 &  0.54-0.76 &    3.62$\pm$0.24 &    5.55$\pm$1.25 &    8.37$\pm$1.42 &   128.6/102 &     d\\
   ngc5466 &  0.00-0.39 &    5.09$\pm$0.66 &   -0.48$\pm$3.41 &    1.48$\pm$1.48 &    31.8/ 39 &     g\\
 &  0.39-0.58 &    5.23$\pm$0.66 &    3.02$\pm$3.37 &    8.37$\pm$4.37 &    24.6/ 38 &     g\\
 &  0.58-0.79 &    4.62$\pm$0.62 &    1.96$\pm$3.37 &    6.41$\pm$4.51 &    34.4/ 39 &     g\\
   ngc5897 &  0.00-0.44 &    4.75$\pm$0.56 &    2.11$\pm$2.32 &    1.98$\pm$0.67 &    41.0/ 63 &     g\\
 &  0.44-0.64 &    4.25$\pm$0.49 &    5.92$\pm$2.29 &    6.65$\pm$0.98 &    29.6/ 59 &     g\\
 &  0.64-0.85 &    5.70$\pm$0.51 &    7.40$\pm$2.29 &   11.82$\pm$1.72 &    32.6/ 61 &     g\\
   ngc5904 &  0.00-0.43 &    4.85$\pm$0.35 &    1.94$\pm$1.49 &    2.35$\pm$1.47 &    75.4/103 &     d\\
 &  0.43-0.68 &    2.10$\pm$0.24 &    6.11$\pm$1.47 &    6.90$\pm$1.40 &    75.9/ 78 &     g\\
 &  0.68-0.99 &    1.15$\pm$0.18 &    4.09$\pm$1.48 &    6.28$\pm$2.12 &    52.3/ 67 &     g\\
   ngc5927 &  0.00-0.75 &    4.69$\pm$0.38 &    4.43$\pm$1.66 &    2.71$\pm$0.86 &   117.0/112 &    df\\
 &  0.75-1.15 &    1.19$\pm$0.34 &   11.01$\pm$1.64 &   15.85$\pm$1.13 &    87.3/ 96 &     f\\
 &  1.15-1.64 &    0.78$\pm$0.39 &   21.00$\pm$1.66 &   35.45$\pm$2.88 &   122.6/ 94 &     f\\
   ngc6093 &  0.00-1.17 &    7.09$\pm$0.44 &    3.29$\pm$1.46 &    3.73$\pm$0.46 &   100.7/108 &     d\\
 &  1.17-1.82 &    2.97$\pm$0.29 &    1.54$\pm$1.46 &    4.25$\pm$0.69 &    89.0/ 85 &     d\\
 &  1.82-2.96 &    0.41$\pm$0.23 &   -0.52$\pm$1.48 &    0.75$\pm$0.96 &    47.2/ 73 &     g\\
   ngc6101 &  0.00-0.84 &    5.74$\pm$0.46 &    3.21$\pm$2.01 &    2.71$\pm$0.81 &    73.0/ 69 &     g\\
 &  0.84-1.25 &    4.76$\pm$0.42 &    7.64$\pm$2.00 &    9.61$\pm$2.16 &    46.1/ 64 &     g\\
 &  1.25-1.72 &    4.21$\pm$0.40 &    6.47$\pm$2.01 &   10.84$\pm$1.38 &    46.2/ 61 &     g\\
   ngc6121 &  0.00-0.20 &    5.49$\pm$0.82 &   -9.74$\pm$4.05 &    0.50$\pm$0.06 &    53.9/ 33 &     n\\
 &  0.20-0.30 &    2.87$\pm$0.80 &   -4.49$\pm$4.04 &    2.47$\pm$2.95 &    29.0/ 35 &     n\\
 &  0.30-0.41 &    4.11$\pm$0.94 &   -8.79$\pm$4.11 &    0.50$\pm$0.00 &    16.4/ 38 &     n\\
       m13 &  0.00-0.50 &    5.00$\pm$0.34 &    1.26$\pm$1.28 &    1.24$\pm$0.43 &   113.3/120 &     d\\
 &  0.50-0.75 &    2.28$\pm$0.26 &    1.67$\pm$1.27 &    2.35$\pm$2.70 &    84.0/102 &     d\\
 &  0.75-1.04 &    1.65$\pm$0.21 &    2.52$\pm$1.27 &    2.71$\pm$0.62 &    86.8/ 90 &     g\\
   ngc6218 &  0.00-0.47 &    3.41$\pm$0.45 &    3.80$\pm$2.35 &    2.35$\pm$0.67 &    51.6/ 59 &     g\\
 &  0.47-0.72 &    2.87$\pm$0.39 &    3.91$\pm$2.35 &    3.70$\pm$1.83 &    39.7/ 55 &     g\\
 &  0.72-1.00 &    3.08$\pm$0.40 &    2.75$\pm$2.36 &    4.44$\pm$2.31 &    54.4/ 57 &     g\\
   ngc6341 &  0.00-0.75 &    5.79$\pm$0.40 &    2.66$\pm$1.24 &    2.35$\pm$0.44 &   113.5/121 &     d\\
 &  0.75-1.18 &    2.77$\pm$0.31 &    3.34$\pm$1.23 &    3.95$\pm$0.43 &    76.3/102 &     d\\
 &  1.18-1.74 &    2.43$\pm$0.25 &    3.69$\pm$1.23 &    4.44$\pm$1.73 &   104.0/ 87 &     g\\
   ngc6352 &  0.00-0.43 &    1.31$\pm$1.24 &   -2.20$\pm$3.50 &    0.50$\pm$0.00 &    36.8/ 57 &     f\\
 &  0.43-0.64 &    1.97$\pm$1.52 &    2.14$\pm$3.77 &    1.48$\pm$3.55 &    36.9/ 54 &     f\\
 &  0.64-0.86 &   -4.64$\pm$1.85 &   -5.00$\pm$4.22 &    0.50$\pm$0.00 &    34.6/ 55 &     f\\
   ngc6362 &  0.00-0.44 &    5.07$\pm$0.57 &    8.48$\pm$2.60 &    4.44$\pm$1.49 &    58.4/ 60 &     g\\
 &  0.44-0.65 &    3.90$\pm$0.52 &    7.41$\pm$2.61 &    7.39$\pm$2.70 &    28.4/ 56 &     g\\
 &  0.65-0.85 &    3.44$\pm$0.49 &   16.05$\pm$2.59 &   29.05$\pm$4.64 &    61.7/ 56 &     g\\
   ngc6397 &  0.00-0.28 &    1.83$\pm$0.75 &   -0.18$\pm$3.27 &    0.50$\pm$0.00 &    36.4/ 46 &     g\\
 &  0.28-0.45 &    1.88$\pm$0.77 &    6.15$\pm$3.28 &   10.84$\pm$1.63 &    33.6/ 45 &     g\\
 &  0.45-0.61 &    4.16$\pm$0.92 &    0.48$\pm$3.36 &   20.19$\pm$7.12 &    43.3/ 49 &     g\\
   ngc6541 &  0.00-0.71 &    4.92$\pm$0.39 &    2.80$\pm$1.39 &    2.59$\pm$0.72 &    97.4/111 &     d\\
 &  0.71-1.14 &    1.52$\pm$0.32 &    2.87$\pm$1.40 &    5.79$\pm$0.79 &    96.3/ 93 &     d\\
 &  1.14-1.66 &    0.86$\pm$0.32 &   11.95$\pm$1.40 &   19.70$\pm$1.41 &   177.0/ 87 &     g\\
   ngc6624 &  0.00-0.91 &    0.93$\pm$0.61 &   12.57$\pm$2.10 &    7.51$\pm$0.68 &    64.2/ 86 &     f\\
 &  0.91-1.45 &    1.51$\pm$0.88 &   14.60$\pm$2.29 &   16.00$\pm$1.11 &    61.0/ 84 &     f\\
 &  1.45-2.13 &    2.15$\pm$1.34 &   10.50$\pm$2.69 &   21.42$\pm$2.71 &    73.2/ 90 &     f\\
   ngc6637 &  0.00-0.87 &    4.90$\pm$0.46 &    2.16$\pm$1.84 &    1.12$\pm$0.83 &   102.9/ 96 &     f\\
 &  0.87-1.38 &    0.87$\pm$0.44 &    5.87$\pm$1.88 &    9.85$\pm$2.96 &    67.2/ 82 &     f\\
 &  1.38-2.15 &   -0.14$\pm$0.65 &    1.80$\pm$2.09 &    1.98$\pm$2.40 &    56.1/ 76 &     f\\
   ngc6652 &  0.00-1.34 &    3.15$\pm$0.66 &    6.04$\pm$3.02 &    3.45$\pm$1.21 &    34.5/ 50 &     f\\
 &  1.34-2.34 &    1.54$\pm$1.06 &    0.27$\pm$3.48 &    4.44$\pm$2.87 &    28.7/ 48 &     f\\
 &  2.34-3.77 &   -4.18$\pm$2.18 &   -9.25$\pm$4.86 &    0.50$\pm$0.00 &    36.2/ 47 &     f\\
   ngc6656 &  0.00-0.26 &   -1.01$\pm$0.47 &   -3.70$\pm$1.73 &    0.50$\pm$0.03 &    86.4/ 84 &     f\\
 &  0.26-0.39 &   -4.94$\pm$0.51 &    1.70$\pm$1.75 &    7.14$\pm$1.02 &   126.0/ 82 &     f\\
 &  0.39-0.52 &   -8.16$\pm$0.58 &   -0.51$\pm$1.82 &    8.37$\pm$1.04 &   168.6/ 80 &     f\\
   ngc6723 &  0.00-0.50 &    6.87$\pm$0.41 &    7.61$\pm$1.72 &    5.67$\pm$0.75 &    93.7/ 92 &     d\\
 &  0.50-0.80 &    3.81$\pm$0.32 &    2.83$\pm$1.74 &    4.31$\pm$1.15 &   115.5/ 79 &     g\\
 &  0.80-1.16 &    2.55$\pm$0.30 &   12.78$\pm$1.72 &   19.14$\pm$1.43 &   132.6/ 72 &     g\\
   ngc6752 &  0.00-0.41 &    0.78$\pm$0.35 &   -1.48$\pm$1.90 &    0.52$\pm$0.03 &    77.4/ 75 &     g\\
 &  0.41-0.66 &   -0.06$\pm$0.26 &    1.30$\pm$1.88 &    0.50$\pm$0.06 &    70.2/ 65 &     g\\
 &  0.66-0.93 &    0.85$\pm$0.25 &    5.24$\pm$1.87 &   11.70$\pm$1.80 &    64.5/ 64 &     g\\
   ngc6809 &  0.00-0.32 &    3.29$\pm$0.45 &    2.00$\pm$2.44 &    1.48$\pm$2.00 &    39.3/ 49 &     g\\
 &  0.32-0.47 &    2.99$\pm$0.46 &    1.85$\pm$2.44 &    4.44$\pm$2.15 &    35.7/ 48 &     g\\
 &  0.47-0.60 &    2.59$\pm$0.40 &    1.01$\pm$2.45 &    3.45$\pm$0.00 &    44.3/ 48 &     g\\
   ngc6981 &  0.00-0.67 &    9.55$\pm$0.58 &    8.62$\pm$2.18 &    5.18$\pm$0.77 &    54.4/ 75 &     d\\
 &  0.67-1.12 &    3.60$\pm$0.40 &    3.55$\pm$2.20 &    3.95$\pm$3.95 &    41.2/ 66 &     g\\
 &  1.12-1.89 &    2.11$\pm$0.32 &    1.14$\pm$2.22 &    2.47$\pm$1.48 &    53.7/ 58 &     g\\
   ngc7078 &  0.00-0.85 &    5.72$\pm$0.40 &    3.97$\pm$1.10 &    2.84$\pm$0.51 &   163.3/140 &    d,e\\
 &  0.85-1.25 &    3.11$\pm$0.31 &    2.45$\pm$1.09 &    2.41$\pm$0.76 &   136.9/122 &    d,e\\
 &  1.25-1.78 &    1.56$\pm$0.25 &   -5.58$\pm$1.12 &    0.50$\pm$0.00 &   252.2/107 &    d,e\\
   ngc7099 &  0.00-0.64 &    5.12$\pm$0.70 &   -4.63$\pm$1.93 &    0.56$\pm$0.46 &    74.6/ 52 &     g\\
 &  0.64-1.11 &    1.39$\pm$0.47 &    3.73$\pm$1.94 &    5.91$\pm$2.15 &    43.0/ 42 &     g\\
 &  1.11-1.74 &    0.67$\pm$0.38 &    2.87$\pm$1.96 &    6.41$\pm$4.29 &    36.0/ 37 &     g\\
      Arp2 &  0.00-0.49 &    6.65$\pm$0.94 &    0.83$\pm$3.85 &    6.41$\pm$2.20 &    13.0/ 28 &     g\\
 &  0.49-0.72 &    7.21$\pm$0.91 &    1.15$\pm$3.89 &   12.31$\pm$12.31 &    11.6/ 29 &     g\\
 &  0.72-1.02 &   10.38$\pm$1.08 &   -7.34$\pm$4.05 &    0.50$\pm$0.00 &     9.1/ 27 &     g\\
 Palomar12 &  0.00-0.40 &   16.60$\pm$2.52 &  -37.87$\pm$8.55 &    0.50$\pm$0.00 &     2.2/ 13 &     n\\
 &  0.40-0.71 &   13.66$\pm$2.15 &  -22.56$\pm$8.15 &    0.50$\pm$0.5 &     1.5/ 13 &     n\\
 &  0.71-1.05 &    9.26$\pm$1.96 &   -8.86$\pm$7.91 &    8.37$\pm$8.37 &     3.6/ 13 &     n\\
   Terzan7 &  0.00-0.86 &   17.38$\pm$1.58 &    6.61$\pm$4.57 &    6.41$\pm$2.20 &    13.0/ 28 &     n\\
 &  0.86-1.48 &    9.85$\pm$1.31 &    3.08$\pm$4.76 &   12.31$\pm$12.31 &    11.6/ 29 &     n\\
 &  1.48-2.35 &    7.94$\pm$1.26 &   -5.75$\pm$5.05 &    0.50$\pm$0.00 &     9.1/ 27 &     n\\
\enddata
\tablecomments{Result flags as in Table ~\ref{tab:whole}.  }
\label{tab:radial2}
\end{deluxetable}

\clearpage

\begin{deluxetable}{ccc}
\tablecolumns{3}
\tablewidth{0pc}
\tablecaption{The core binary fractions for 25 clusters with the high q method}
\tablehead{
Name &  fb(high q)\% & Note$^a$
}
\startdata
ngc104 & 4.26 $\pm$ 2.68 &d,n\\
ngc288 & 5.64 $\pm$ 0.41 &g\\
ngc4590 & 9.43 $\pm$ 0.69 &g\\
ngc5053 & 5.67 $\pm$ 0.40 &g\\
m3 & 5.42 $\pm$ 1.07 &d\\
ngc5466 & 5.58 $\pm$ 0.43 &g\\
ngc5897 & 5.08 $\pm$ 0.34 &g\\
ngc5904 & 7.73 $\pm$ 0.72 &d\\
ngc5927 & 5.41 $\pm$ 0.82 &d,f\\
ngc6101 & 6.13 $\pm$ 0.40 &g\\
ngc6121 & 4.94 $\pm$ 0.63 &g\\
m13 & 6.17 $\pm$ 0.47 &d\\
ngc6218 & 3.75 $\pm$ 0.47 &g\\
ngc6341 & 1.20 $\pm$ 2.47 &d,n\\
ngc6352 & 1.76 $\pm$ 1.30 &f\\
ngc6362 & 4.81 $\pm$ 0.44 &g\\
ngc6541 & -0.12 $\pm$ 5.47 &d,n\\
ngc6637 & 8.51 $\pm$ 1.27 &d,f\\
ngc6656 & -2.39 $\pm$ 0.33&f \\
ngc6723 & 6.89 $\pm$ 0.38 &g\\
ngc6752 & 5.27 $\pm$ 2.47&d,n \\
ngc6809 & 3.43 $\pm$ 0.24&g \\
ngc6981 & 11.56 $\pm$ 0.81&g \\
Arp2 & 7.42 $\pm$ 0.69 &n\\
Terzan7 & 20.02 $\pm$ 2.10 &n\\
\enddata
\tablecomments{Result flags as in Table ~\ref{tab:whole}.}
\label{tab:coreFb}
\end{deluxetable}
\clearpage

\begin{deluxetable}{cccccc}
\tablecolumns{6}
\tablewidth{0pc}
\tablecaption{Fitting results for binary fractions vs. radius}
\tablehead{
normalize &  k & b & $\chi^2/dof$ & coefficient$_r$ & significance
}
\startdata
to $r_c$ and $fb_c$&  -0.41$\pm$0.04& $1.34\pm0.07$ & 286.5/64& -0.69  & 1.9e-10\\
to $r_h$ and $fb_h$& -0.99$\pm$0.08 & $1.45\pm0.06$ & 288.3/100& -0.61 & 7.4e-12\\
\enddata
\label{tab:Fb_r}
\end{deluxetable}

\clearpage

\begin{figure}
\plottwo {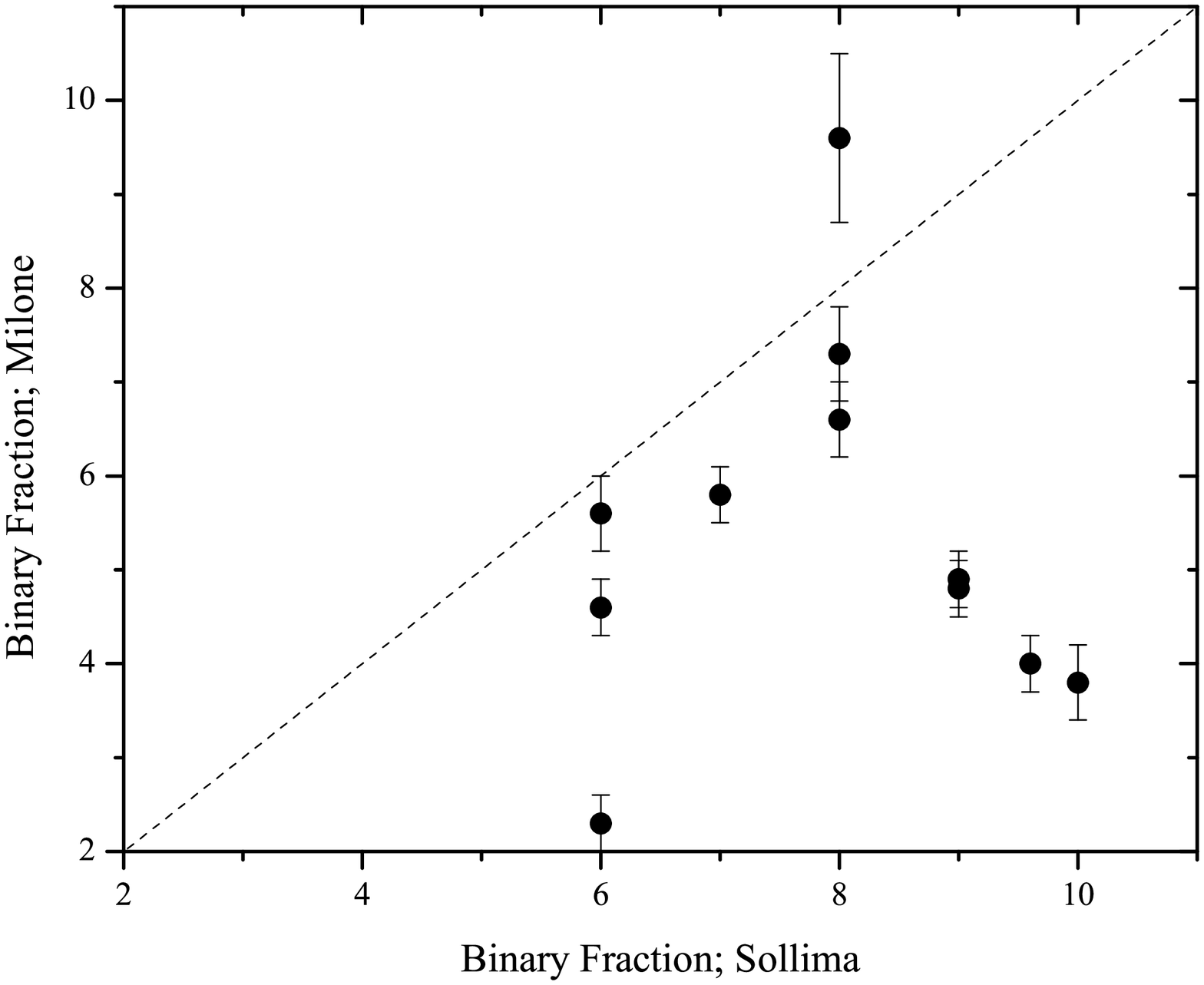} {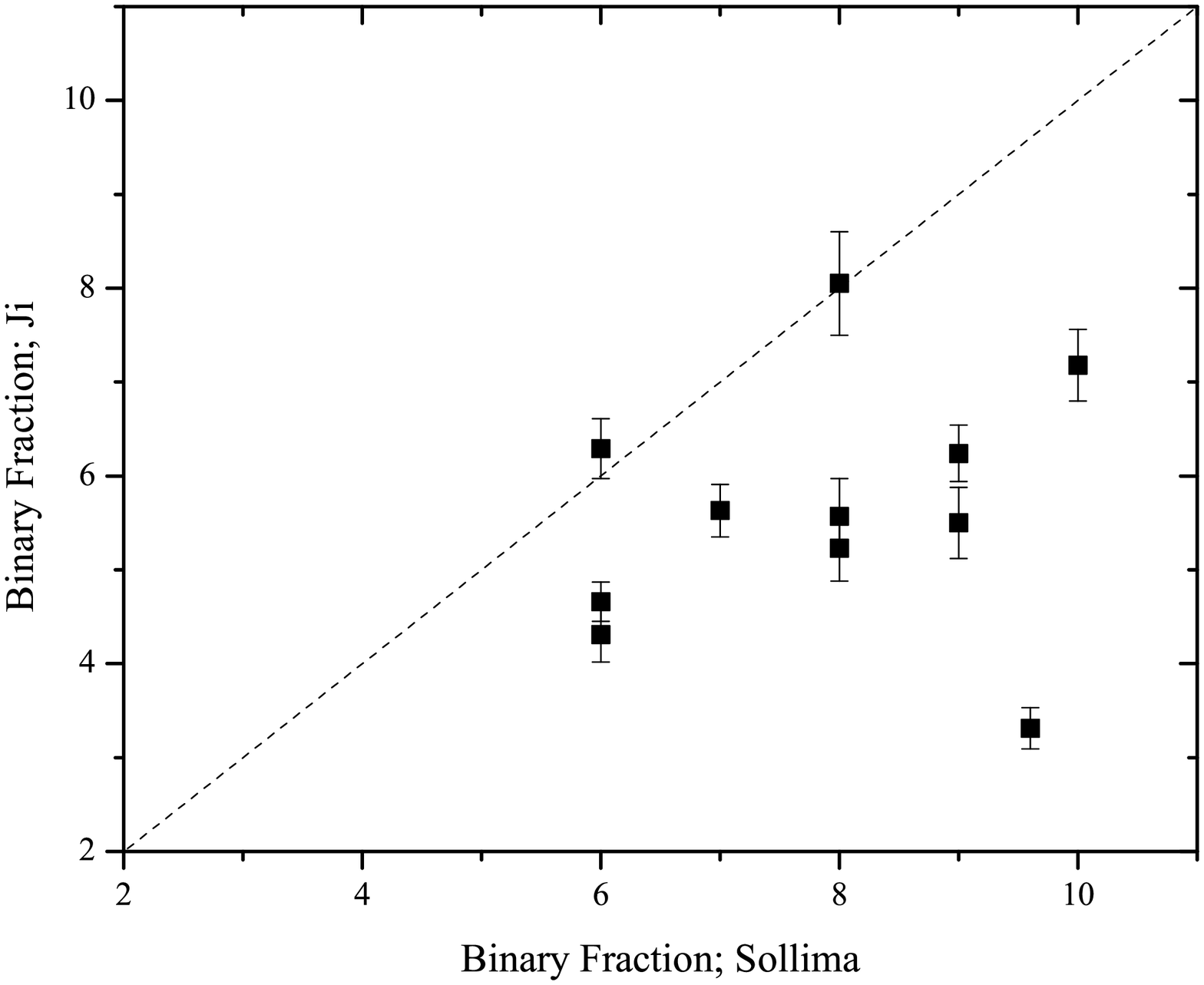}
\caption{A comparison between the whole field binary fractions obtained by \citet{sollima07} and that of \citet{milone12} is shown on the left, where the estimated errors by Sollima is about 1\%.  
The two clusters with the largest binary fractions, near 20\% (Arp 2 and Terzan 7), are not shown here.
The dotted line is where the points would like if there were a perfect correlation.  The relationship between the two samples is poor.
When comparing the values from \citet{sollima07} to those presented in this work (right panel), the relationship is still poor, with the values of Sollima being systematically higher.}
\label{fig:comp_Solima}
\end{figure}


\begin{figure}
\plotone {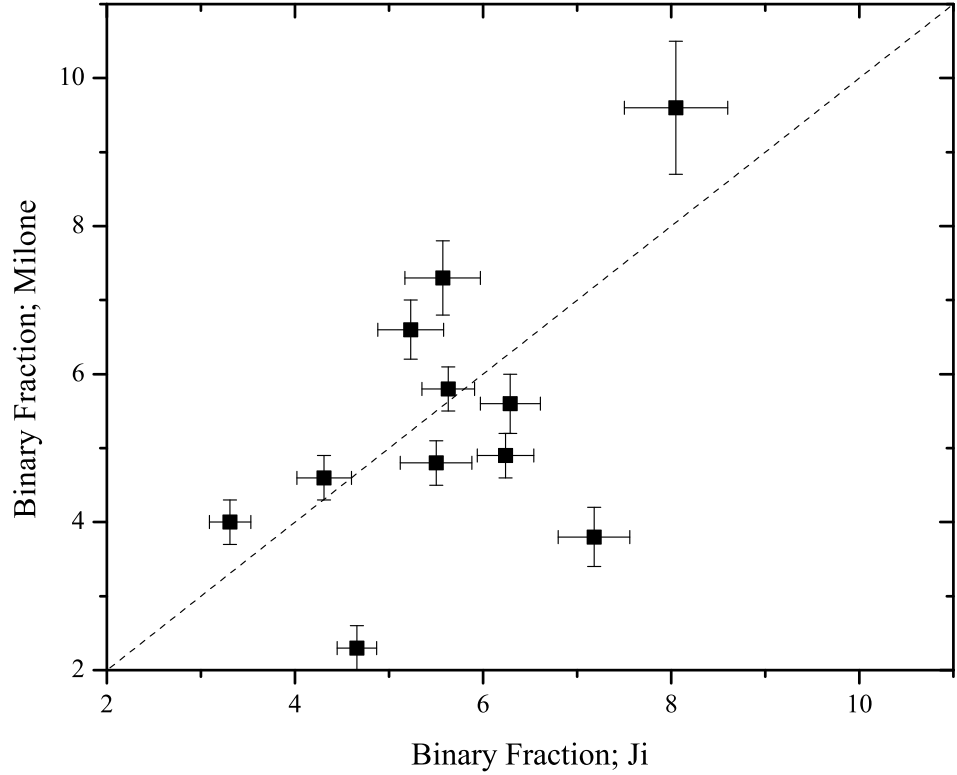}
\caption{A comparison between the whole field binary fractions obtained by \citet{milone12} and
this work (excluding Arp 2 and Terzan 7), for the clusters used by \citet{sollima07}.  While there is a general correlation between the two sets of values, there are two significant discrepancies.
}
\label{fig:comp_Mil_Ji}
\end{figure}


\begin{figure}
\plottwo{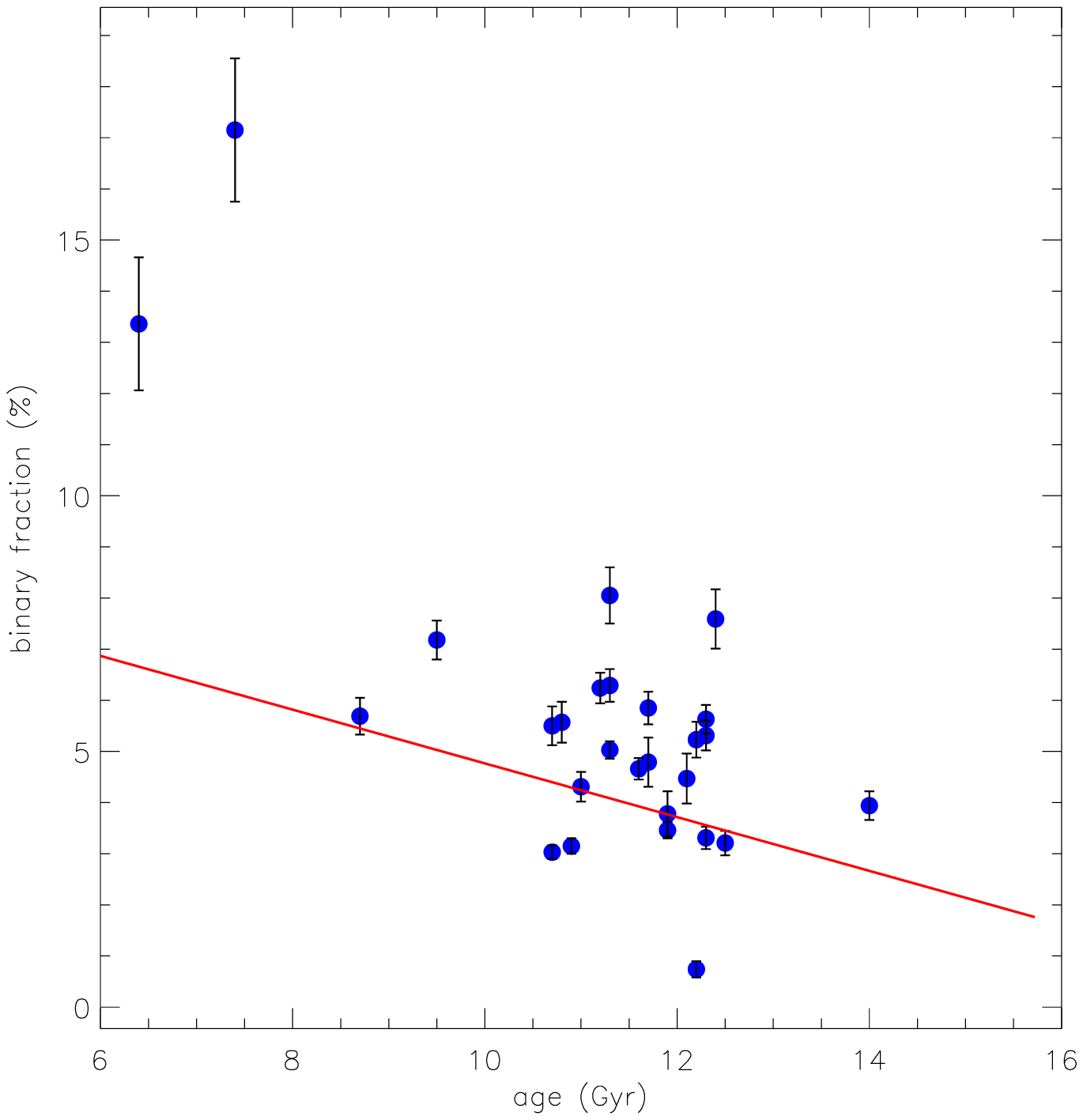} {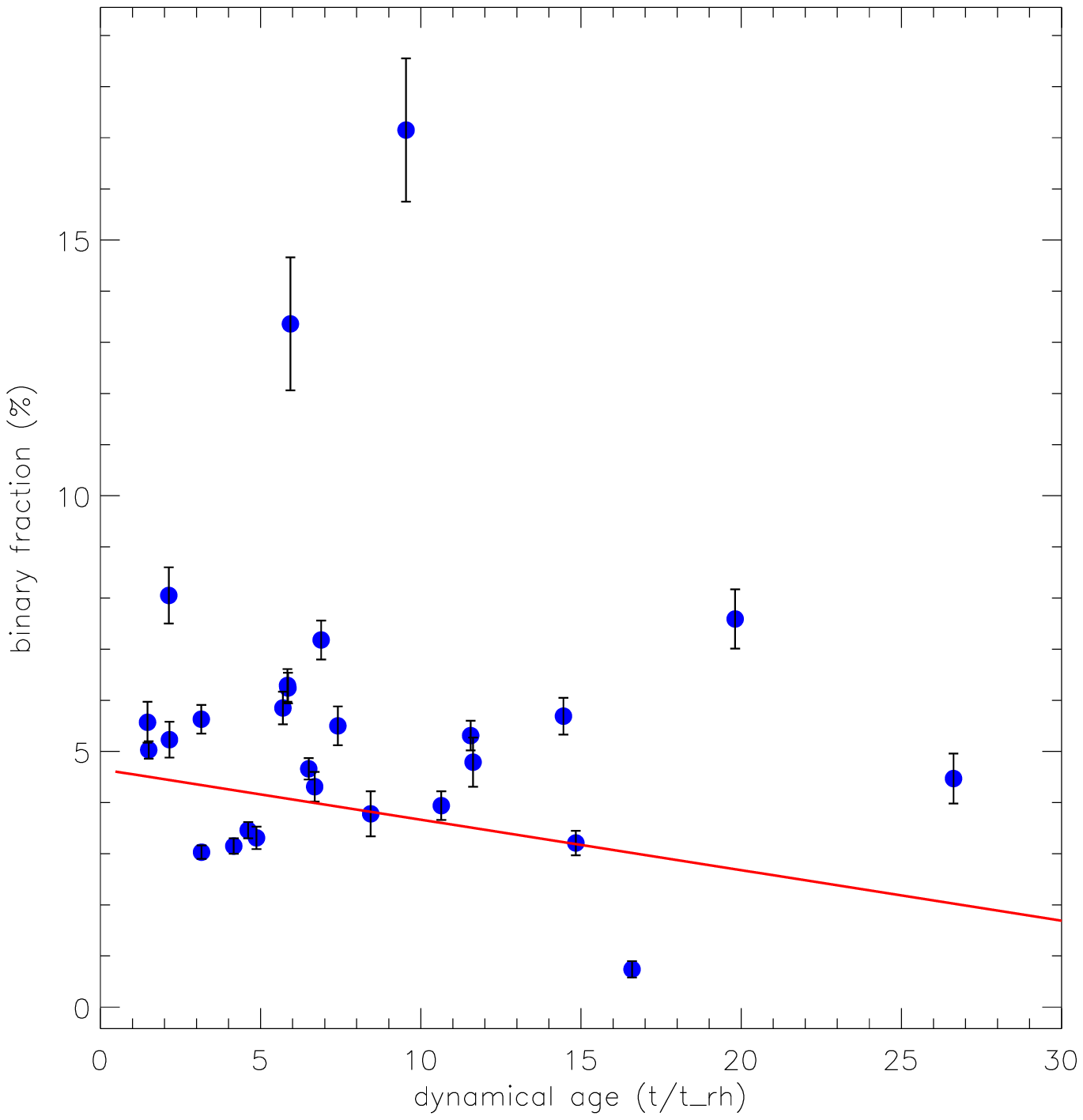}
\plottwo{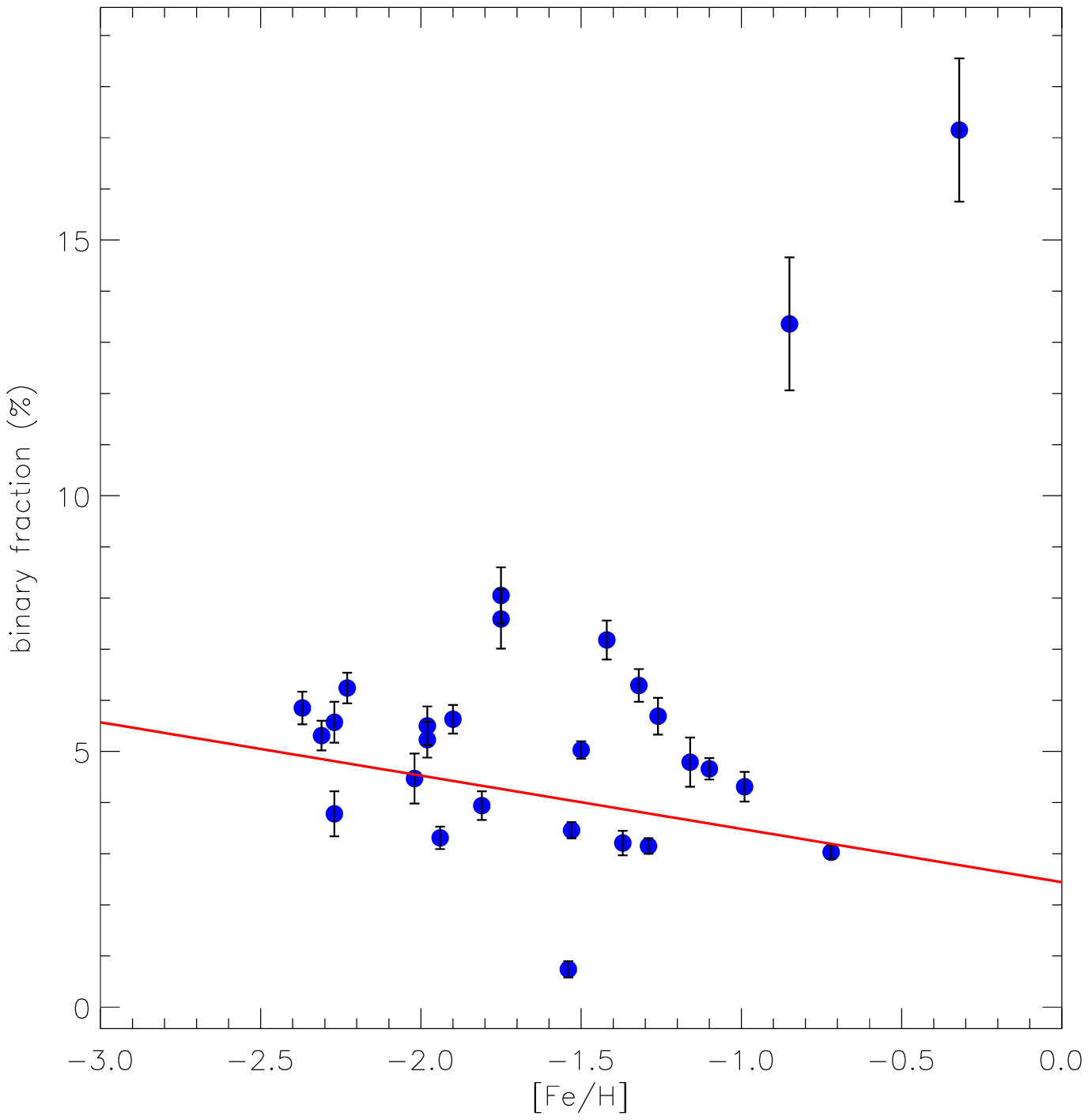} {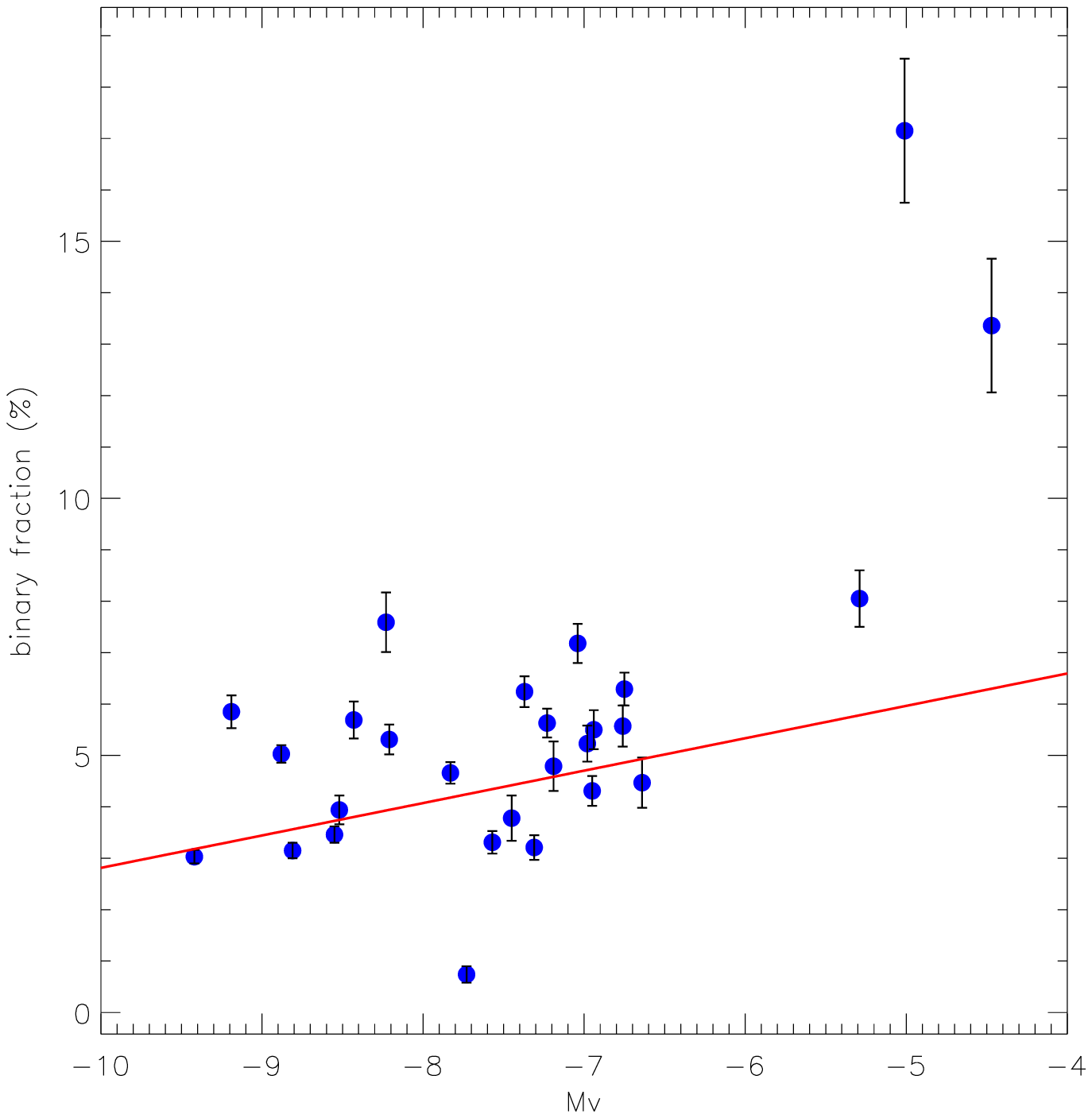}
\caption{The half-mass radius binary fractions as a function of different cluster properties. Upper left: fb(half-mass radius) vs ages; upper right: fb vs dynamical ages; Lower left: fb vs [Fe/H]; Lower right: fb vs Mv.}
\label{fig:rh_Fb}
\end{figure}

\begin{figure}
\plotone {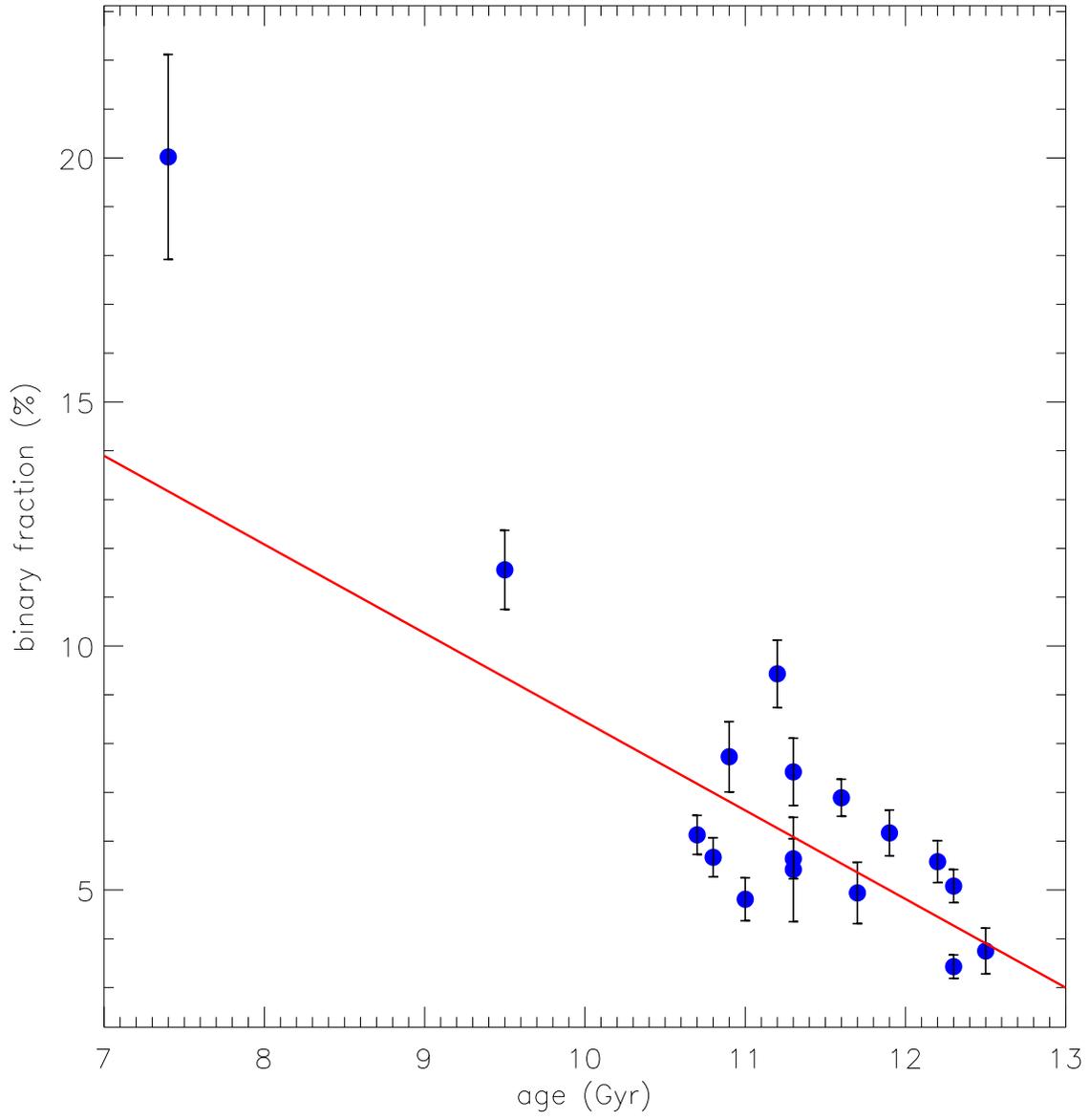}
\caption{Our core binary fractions has a significant dependence on the cluster age and the relationship would be significant even if the two youngest clusters were removed.}
\label{fig:coreFb}
\end{figure}

\begin{figure}
\plotone {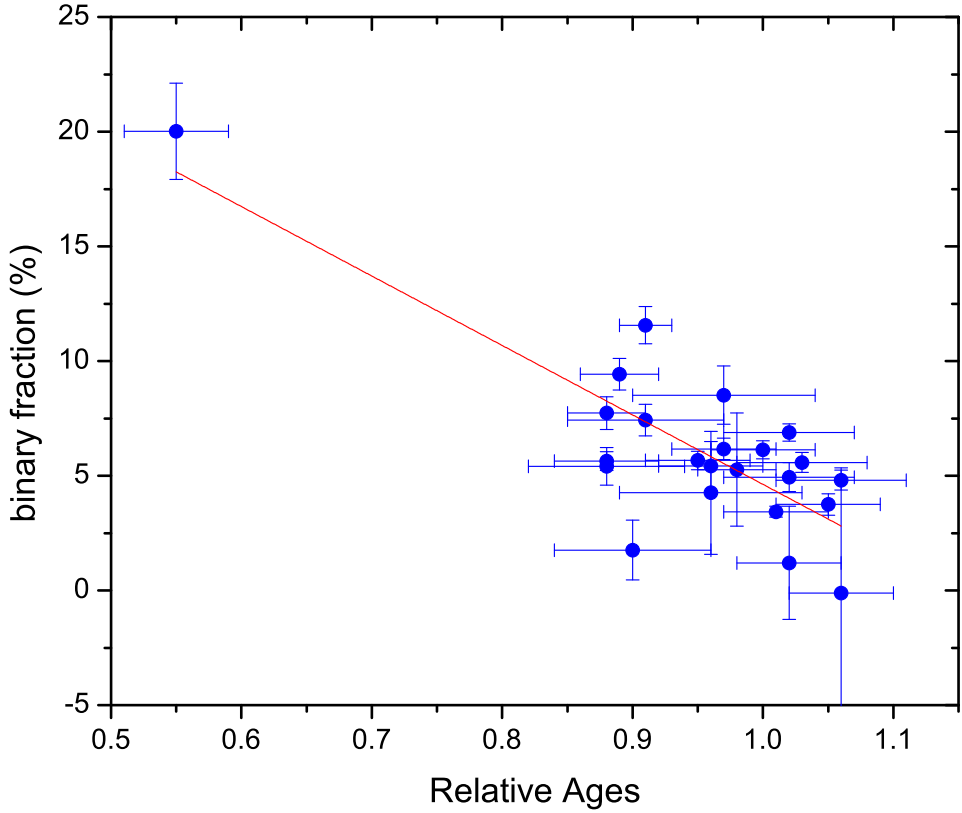}
\caption{Similar to Figure 5, except relative ages from \citet{marin09} are used and we included several clusters with large numbers of foreground/background stars, which leads to negative binary fractions with large uncertainties.  The relationship still exists whether or not these are included.  A anticorrelation between binary fraction and age is confirmed when using this independent set of age measurements.}
\label{fig:coreFb2}
\end{figure}

\begin{figure}\centering
\plottwo {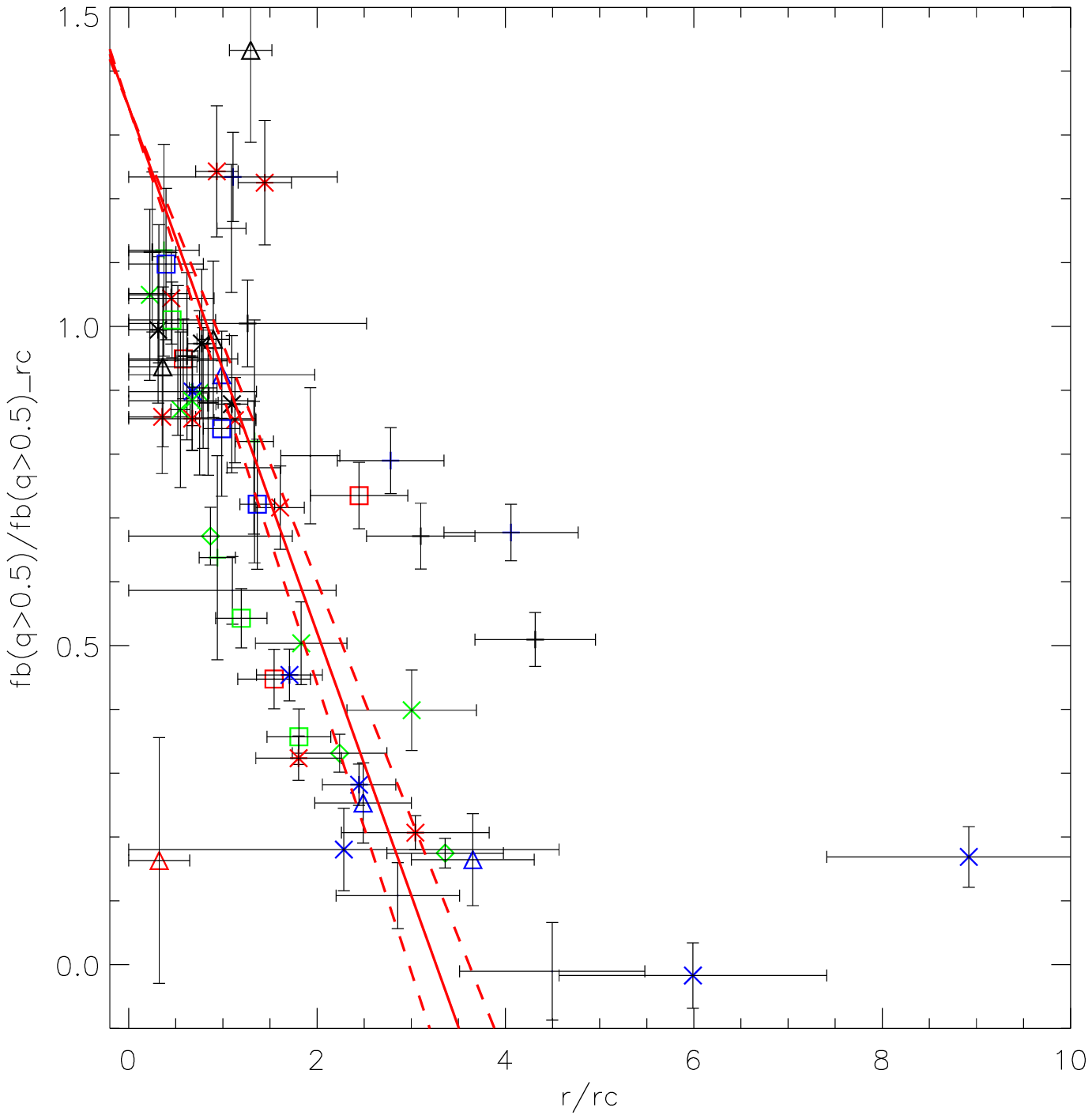} {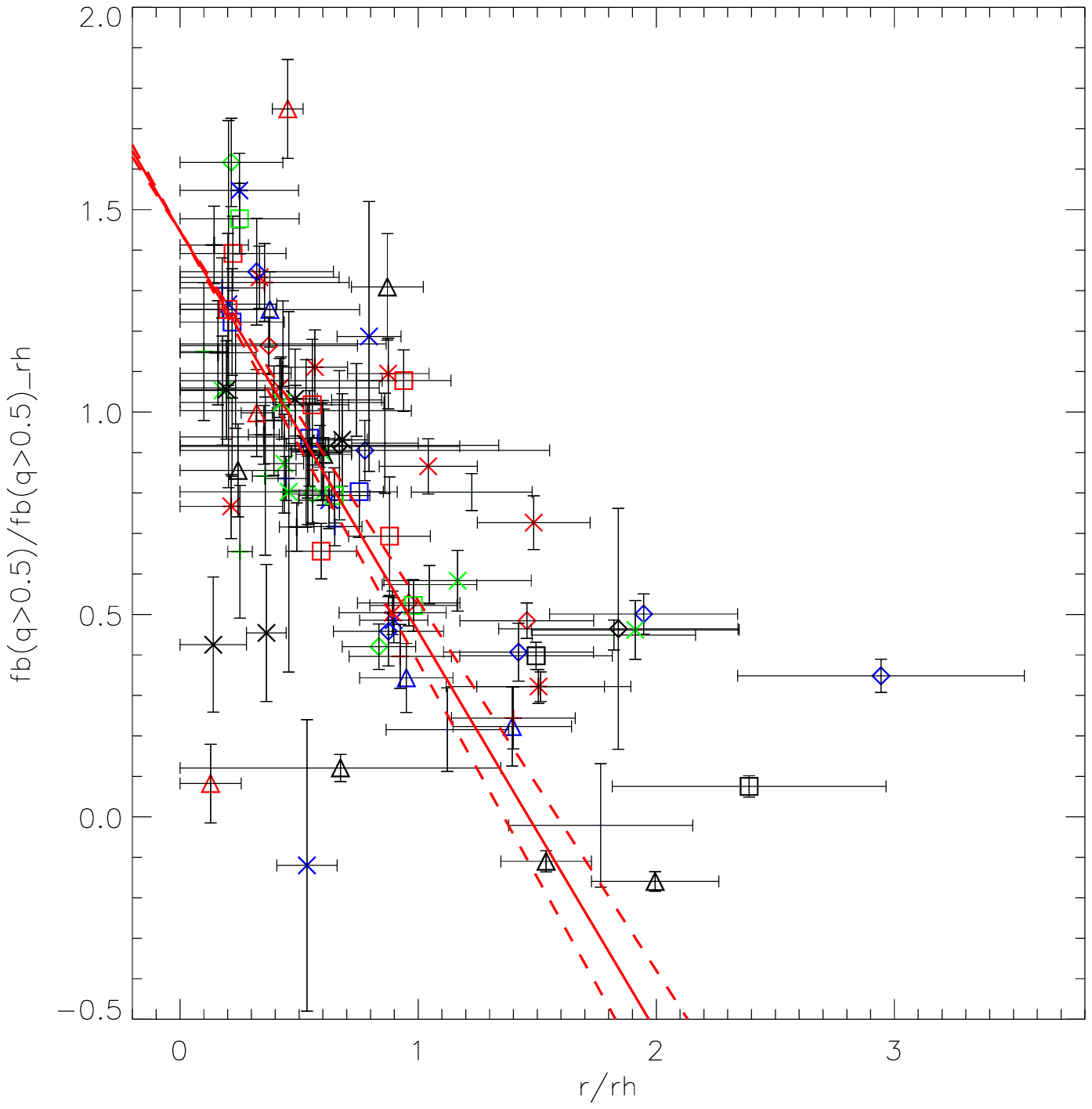}
\caption{Combined high mass-ratio ($q>0.5$) binary fractions as a function of radius. Each cluster has three bins, and with the same symbol and the same color on this figure.  This strong relationship is similar to that found by \citet{milone12}.}
\label{fig:Fb_r}
\end{figure}

\begin{figure}\centering
\plotone {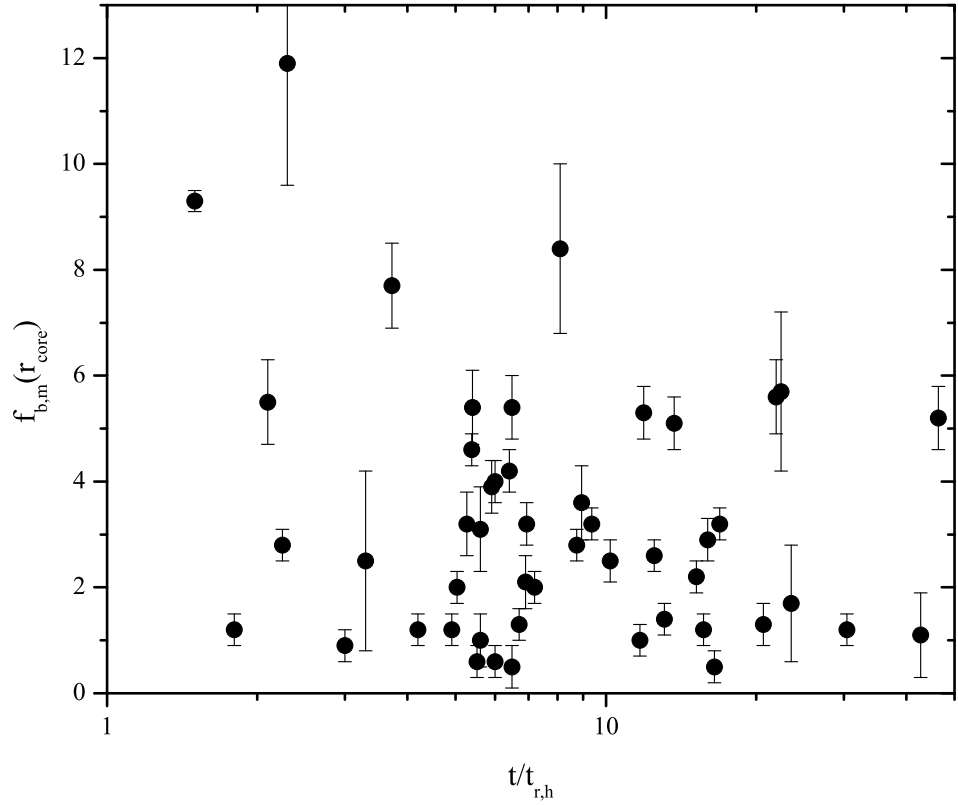}
\caption{The distribution of binary fractions within the core, determined by \citet{milone12}, 
as a function of the dynamical age, being the cluster age divided by the relaxation time at the 
half-light radius.  The dominant feature is the nearly order of magnitude variation in the binary 
fraction for a given dynamical age, even at early dynamical times.  This suggests that clusters 
were born with a wide range of binary fractions.  The data suggest a modest decrease of binary fraction with dynamical time.
}
\label{fig:binary_dyn_t}
\end{figure}

\end{document}